\documentclass[showpacs,amsfonts,aps,prc,floatfix,notitlepage,%
superscriptaddress,nofootinbib,10pt,showkeys]{revtex4-1}
\usepackage{amsmath}
\usepackage{bm}
\usepackage{graphicx}
\usepackage{url}
\usepackage{longtable}
\usepackage{bbm}
\usepackage{hyperref}
\allowdisplaybreaks

\usepackage{color}
\def\vv\label#1{%
  \label{#1}\relax%
  \makebox[0.1em]{%
    \raisebox{1.1em}[0em][0em]{%
      \scriptsize\textcolor[rgb]{0.0,0.0,1.0}{%
        \textbackslash{ref}\{#1\}}}}}
\def\vv\label#1{\label{#1}}

\newcommand{\eqdef}{:=}

\newcommand{\thatis}{{\it i.e.}}
\newcommand{\D}{\mathrm{D}}
\renewcommand{\d}{d}
\newcommand{\ud}[2]{^{#1}{}_{#2}}
\newcommand{\du}[2]{_{#1}{}^{#2}}
\DeclareMathOperator*{\diag}{diag}
\DeclareMathOperator*{\const}{const}

\newcommand{\identity}{\mathbbm{1}}
\newcommand{\T}{\mathrm{T}}

\def\ud#1#2{^{#1}{}_{#2}} % ordered index
\newcommand{\eqsym}{\stackrel{\textrm{sym}}{=}}

\def\[#1\]{\begin{align*}#1\end{align*}}
\def\(#1\){\begin{align}#1\end{align}}

\begin{document}

%%%%%%%%%%%%%%%%%%%%%%%%%%%%%%%%%%%%%%%%%%%%%%%%%%%%%%%%%%%%%%%%%%%%%%%%%%%%%%%

\title{Causal hydrodynamic fluctuations
  in non-static and inhomogeneous backgrounds}
\date{\today}

\author{Koichi Murase}
\email{murase@nt.phys.s.u-tokyo.ac.jp}
\affiliation{Department of Physics, Sophia University,
  7-1 Kioicho, Chiyoda-ku, Tokyo 102-8554, Japan}
\affiliation{Department of Physics, The University of Tokyo,
  7-3-1 Hongo, Bunkyo-ku, Tokyo 113-0033, Japan}

\begin{abstract}
To integrate hydrodynamic fluctuations, namely thermal fluctuations of hydrodynamics,
into dynamical models of high-energy nuclear collisions based on relativistic hydrodynamics,
the property of the hydrodynamic fluctuations
given by the fluctuation--dissipation relation should be carefully investigated.
The fluctuation--dissipation relation
for causal dissipative hydrodynamics with the finite relaxation time
is naturally given in the integral form of the constitutive equation
by the linear-response theory.
While, the differential form of the constitutive equation
is commonly used in analytic investigations and dynamical calculations
for practical reasons.
We give the fluctuation--dissipation relation
for the general linear-response differential form
and discuss the restrictions to the structure of the differential form,
which comes from the causality and the positive semi-definiteness of the noise autocorrelation,
and also the relation of those restrictions to the cutoff scale of the hydrodynamic fluctuations.
We also give the fluctuation--dissipation relation for the integral form
in non-static and inhomogeneous background
by introducing new tensors, the pathline projectors.
We find new modification terms to the fluctuation--dissipation relation
for the differential form in non-static and inhomogeneous background
which are particularly important in dynamical models to describe rapidly expanding systems.
\end{abstract}

%%\pacs{25.75.-q, 25.75.Nq, 12.38.Mh, 12.38.Qk}

\maketitle
\tableofcontents

\section{Introduction}
Relativistic hydrodynamics has been widely used in various fields
such as cosmology, astrophysics and nuclear physics
to describe spacetime evolution of thermodynamic fields
in a vast range of the scales from femtometer scale to astronomical scales.
In this paper we argue general properties of thermal fluctuations
of relativistic hydrodynamics
which are originally needed to investigate
the effects of the fluctuations on experimental observables of high-energy nuclear collisions
but can also be applied to any relativistic systems
with the situations the thermal fluctuations have relevant effects.

The aim of the high-energy nuclear collision experiments
is to create and understand extremely hot and/or dense nuclear matter
whose fundamental degrees of freedom are quarks and gluons.
The dynamics of systems of quarks and gluons are
ruled by quantum chromodynamics (QCD).
While quarks and gluons are confined in hadrons
in normal temperature and density,
they are deconfined to form a novel state of the matter
called quark gluon plasma (QGP)~\cite{Baym:1976yu,Kapusta:1979fh,Shuryak:1977ut,Yagi:2005yb}
in extremely hot and/or dense circumstances
such as in the early universe at microseconds after the big bang.
The QGP can be experimentally created in high-energy nuclear collision experiments
at Relativistic Heavy Ion Collider (RHIC) in Brookhaven National Laboratory
and Large Hadron Collider (LHC) in CERN
to study the properties of the created QGP\@.

The key to successfully extract the properties of the created QGP from experimental data
is proper dynamical modeling of the collision reaction.
The direct observable in the experiments
is the distribution of final-state hadrons
which formed after the non-equilibrium spacetime evolution of the created QGP\@.
To reconstruct from the hadron distribution
the information on the non-equilibrium dynamics of the created matter
and to determine the transport properties of the QGP,
sophisticated dynamical models,
which describe the whole process of the high-energy nuclear collision reaction
from the initial state and the QGP to the final state hadron gases,
are needed. The standard structure of modern dynamical models
is the combination of
initial state models which describe the initial thermalization process,
relativistic hydrodynamics which follows the spacetime evolution of the created matter,
and hadronic transport models which describe the final-state interactions.
Among them, the most important part is relativistic hydrodynamics
reflecting the properties of the QGP\@.
In the early 2000s, dynamical models based on relativistic ideal hydrodynamics~\cite{Kolb:2000sd,Teaney:2000cw,Huovinen:2001cy,Hirano:2002ds}
has been successful to reproduce the experimental observables from RHIC
such as the elliptic flows $v_2$~\cite{Adler:2001nb,*Adler:2002pu,Adcox:2002ms,*Adler:2003kt,Back:2002gz,*Back:2004zg,*Back:2004mh},
and the new paradigm of the strongly-coupled QGP (sQGP)
with tiny viscosity has been established~\cite{Heinz:2000bk,*Heinz:2001xi,Gyulassy:2004vg,*Gyulassy:2004zy,Muller:2006ee}.
Since then, the hydrodynamic part of dynamical models
has been updated to include dissipation such as shear viscosity and bulk viscosity
and, along with sophisticated initial conditions and hadronic transport models,
used to determine the quantitative values of such transport coefficients%
~\cite{Luzum:2008cw,Schenke:2010rr,Song:2010mg,Bernhard:2016tnd}.

One of the important physics in determining such transport properties of the created matter
is event-by-event fluctuations of collision reactions.
For example, the flow harmonics $v_n$
reflects the anisotropy of the created matter 
caused by event-by-event fluctuations of collision reactions.
The major part of the higher-order flow harmonics
is determined by the initial-state fluctuations
of the distributions of nucleons in colliding nuclei.
Nevertheless other different kinds of fluctuations,
such as thermal fluctuations and jets,
have non-negligible contributions to the flow observables~\cite{Kapusta:2011gt}.
Another example is the event-by-event fluctuations
of the conserved charges which could be used
as a signal of the critical point
and the first-order phase transition of the QGP
created in lower-energy collisions.
For the search of the QCD critical point in high-baryon density domain of the phase diagram,
a number of experiments such as the Beam Energy Scan programs at RHIC,
NA61/SHINE experiment at SPS,
CBM experiment at FAIR, MPD at NICA, CEE at HIAF
and the heavy-ion program at J-PARC are ongoing or planned.
To quantitatively determine the properties of the created matter,
and to find signals of a critical point
in the finite baryon region of the QCD phase diagram,
it is indispensable to investigate the physical properties of these fluctuations
and properly implement them in dynamical models.

Hydrodynamic fluctuations~\cite{LandauLifshitz:FluildMechanics,LifshitzPitaevskii:StatisticalPhysics},
\thatis, the thermal fluctuations of hydrodynamics,
is one of the sources of event-by-event fluctuations of high-energy nuclear collisions.
Also, the hydrodynamic fluctuations near the critical point
play an important role in the dynamics
near the critical point and the first-order phase transition,
so the dynamical description of the hydrodynamic fluctuations
is becoming increasingly important.
Hydrodynamic fluctuations are introduced in hydrodynamic equations as noise fields,
and the noise power spectrum is determined by viscosity and diffusion coefficients
through the fluctuation--dissipation relation (FDR)~\cite{Kubo:1957mj}.
Now the hydrodynamic equations become stochastic partial differential equation,
and such a framework is called fluctuating hydrodynamics~\cite{LandauLifshitz:FluildMechanics,LifshitzPitaevskii:StatisticalPhysics}.
Relativistic fluctuating hydrodynamics
is first considered in the first-order theory~\cite{Calzetta:1997aj,Kapusta:2011gt},
and then extended for the second-order theory to apply it to the high-energy nuclear collisions~\cite{Murase:2013tma,Young:2013fka}.
The effects of hydrodynamic fluctuations to the observables in dynamical models
are first investigated in linearized fluctuating hydrodynamics~\cite{Young:2014pka}
and then in fully non-linear relativistic fluctuating hydrodynamics~\cite{Murase:2016rhl,Nagai:2016wyx,Sakai:2017rfi,Singh:2018dpk,Sakai:2018sxp}.
For the critical point search, there are already several simulations
solving the hydrodynamic fluctuations in the baryon current or other slow modes in simple setups%
~\cite{Sakaida:2017rtj,Nahrgang:2017oqp,Bluhm:2018plm,Nahrgang:2018afz,Herold:2018ptm,Jiang:2017mji,Wu:2017dyu,Wu:2018twy,Wu:2019qfz}.
Instead of the Langevin type description (``event-by-event'' description) of fluctuations,
also the extension of the hydrodynamics (Hydro+) with slow modes and two-point functions is proposed for the critical dynamics~\cite{Stephanov:2017ghc}.
Another important topic of the hydrodynamic fluctuations
is the renormalization of hydrodynamics
due to the non-linear effects of the hydrodynamic fluctuations
which are analyzed by various methods in various contexts%
~\cite{Kovtun:2003vj,Kovtun:2011np,PeraltaRamos:2011es,Akamatsu:2016llw,Akamatsua:2017yph,Akamatsu:2017rdu,Hongo:2018cle,Martinez:2018wia,An:2019osr}.
The transport coefficients and the equation of state should be renormalized
depending on the cutoff scales of the hydrodynamic fluctuations,
and also there arises the long-time tail of the two-point correlations,
which cannot be renormalized into ordinary transport coefficients.
The renormalization of the transport coefficients and the equation of state
has already turned out to be important in dynamical models
of the high-energy nuclear collisions~\cite{Murase:2016rhl,Nahrgang:2017oqp}.
The proper choice of the cutoff in dynamical models
is also an important problem,
for which the theoretical estimation of the bound of the cutoff
has been given in the context of renormalization~\cite{Kovtun:2014nsa,Martinez:2017jjf}.
In existing analyses and calculations,
the first-order hydrodynamics is usually assumed,
or, even if the second-order causal hydrodynamics is assumed,
a naive expression for the FDR in the global equilibrium has been used.
However, the matter created in the high-energy nuclear collisions
is rapidly expanding and highly inhomogeneous,
so that the FDR should be carefully reconsidered.

In this paper we consider the properties of hydrodynamic fluctuations
in causal dissipative hydrodynamics
to clarify a proper treatment of the hydrodynamic fluctuations
of non-static and inhomogeneous matter
in dynamical models which is consistent with the causality.
In the first-order dissipative hydrodynamics,
which is also known as the Navier--Stokes theory,
the hydrodynamic fluctuations are the white noise according to the FDR,
\thatis, the autocorrelation of the hydrodynamic fluctuations is a delta function in space and time.
However, the Navier--Stokes theory has
problems of acausal propagation of information and unstable modes~\cite{Hiscock:1983zz,*Hiscock:1985zz,*Hiscock:1987zz}.
Instead, in dynamical models the second-order dissipative hydrodynamics is commonly used
because the problem of the acausality is known to be resolved in the second-order hydrodynamics
with an appropriate value of the relaxation time~\cite{Israel:1976tn,Israel:1979wp}.
In the second-order causal dissipative hydrodynamics,
the dissipative currents are treated as dynamical fields,
and they respond to the thermodynamic forces with non-zero time of the relaxation time scale.
To respect the causality,
the relaxation time of the dissipative currents has a lower bound,
which can be understood naively in the following way:
the relaxation is caused by the interaction with many degrees of freedom in the system,
so a non-zero time span is needed to achieve a sufficient interaction for the relaxation
because the interaction is restricted within the light cone in relativistic systems.
In such relativistic systems, the hydrodynamic fluctuations of the dissipative currents
should also have finite relaxation time.
The autocorrelation of the hydrodynamic fluctuations is no longer the delta function
but has non-zero values within the time scale of the relaxation time~\cite{Murase:2013tma,Young:2013fka},
which means that the FDR should be modified to a colored one from the white one of the Navier--Stokes theory.
In the third--order theory, the noise has non-zero autocorrelations also in a finite spatial range~\cite{Kapusta:2014dja}.

In addition, another subtlety of the FDR arises
with the non-zero relaxation time in non-static and inhomogeneous systems.
Usually, the FDR is obtained
by considering the linear response of the global equilibrium state to small perturbations,
but it is non-trivial how the FDR is modified in non-static and inhomogeneous backgrounds.
Nevertheless,
in the case of the Navier--Stokes theory
the FDR is a delta function so that
it can be applied to solve the dynamics using the local temperature and chemical potentials
even in non-static and inhomogeneous backgrounds.
However, in the case of the second-order causal theory,
the FDR is non-local so that a single thermal state cannot be assumed
to determine the temperature and chemical potential
appearing in the FDR of the global equilibrium.
Here we need to explicitly consider the FDR in non-static and inhomogeneous systems.
First, in Sec.~\ref{sec:hydro}, the integral form
and the differential form of the constitutive equation
are introduced to discuss the relation between the integral form in the linear-response regime
and the differential form used in actual dynamical simulations.
In Sec.~\ref{sec:pathline}, to deal with the tensor structure of the integral form
for the shear stress tensor and the diffusion currents
in non-static and inhomogeneous backgrounds,
new tensors, which are called the pathline projectors in this paper, are introduced.
Next, in Sec.~\ref{sec:fdr}, we obtain the FDR in the causal theories
and discuss the restriction on the structure of the general linear-response differential forms
and its relation to the cutoff scales of the hydrodynamic fluctuations.
Finally, in Sec.~\ref{sec:inhomogeneous}, the FDR in non-static and inhomogeneous backgrounds
are obtained for the integral and differential forms using the pathline projectors.
Section~\ref{sec:summary} is devoted for the summary of the conclusions and the discussions.
Hereafter, we adopt the natural unit system with $c = k_B = 1$ and
the metric tensor with the sign convention being $g_{\mu\nu} = \diag(1, -1, -1, -1)$.

\section{Constitutive equation}
\label{sec:hydro}
In this section we review the constitutive equation in relativistic hydrodynamics
and introduce its two representations
which are used to discuss
the higher-order theory and the causality~\cite{Koide:2006ef}.

Hydrodynamics is based on the conservation laws:
\begin{align}
  \partial_\mu T^{\mu\nu} &= 0,
    \vv\label{eq:hydro.conservation-T} \\
  \partial_\mu N_i^{\mu} &= 0, \quad (i=1,\dots,n),
    \vv\label{eq:hydro.conservation-N}
\end{align}
where the conserved currents, $T^{\mu\nu}$ and $N_i^{\mu}$, are
the energy-momentum tensor and the Noether current for the $i$-th conserved charge, respectively.
In the Landau frame in which the flow vector $u^\mu$ is defined to be
the timelike eigenvector of $T^\mu{}_\nu$ normalized as $u_\mu u^\mu = 1$,
the conserved currents are decomposed
into irreducible tensor components of SO(3) rotation in the local rest frame of the matter as
\begin{align}
  T^{\mu\nu} &= e u^\mu u^\nu - (P+\Pi) \Delta^{\mu\nu} + \pi^{\mu\nu}, \\
  N_i^{\mu} &= n_i u^\mu + \nu_i^\mu,
\end{align}
where $\Delta^{\mu\nu} \eqdef g^{\mu\nu} - u^\mu u^\nu$
is the projector for vectors onto the spatial components in the local rest frame.
The local conserved quantities, $(e, n_i)$, are the local energy density
and the local $i$-th charge density, respectively.
The symbol $P$ denotes the equilibrium pressure.
The dissipative currents, $(\Pi, \pi^{\mu\nu}, \nu_i^\mu)$,
are the bulk pressure, the shear-stress tensor and the diffusion current for the $i$-th charge,
respectively.
The hydrodynamic equations can be written
in terms of the flow velocity field $u^\mu(x)$ (for which the number of degrees of freedom is $\#d.o.f.=3$),
the thermodynamic fields, $(e(x), n_i(x), P(x))$ ($\#d.o.f. = 1 + n + 1$),
and the dissipative currents $(\Pi(x), \pi^{\mu\nu}(x), \nu_i^\mu(x))$ ($\#d.o.f. = 1 + 5 + 3n$).
The number of degrees of freedom is $11+4n$ in total.
The conservation laws, \eqref{eq:hydro.conservation-T} and \eqref{eq:hydro.conservation-N},
give $(4+n)$ dynamical constraints.
The equation of state, $P(x)=P(e(x),n_i(x))$, gives another constraint.
The rest $(6+3n)$ degrees of freedom are constrained by the constitutive equations
describing the dissipative currents $\Gamma(x) \eqdef (\Pi(x), \pi^{\mu\nu}(x), \nu_i^\mu(x))^\T$.
The dissipative currents, $\Gamma(x)$, can be in general written as
the functional of the present ($x'^0=x^0$) and past ($x'^0 < x^0$)
dynamical fields, $X(x') \eqdef (u^\mu(x'), e(x'), n_i(x'), \Gamma(x'))$:
\begin{align}
  \Gamma(x) &= \Gamma[\{X(x')\}_{x'^0 \le x^0}](x).
\end{align}
However, this form is too general
to theoretically or experimentally determine the explicit behavior of the dissipative currents.
Using various conditions and approximations,
such as symmetry, the second law of thermodynamics,
the gradient expansion or the linear-response theory,
we shall obtain more restricted forms to express the behavior of the dissipative currents.

\subsection{Differential form of constitutive equation}
If one ignores treatment of the convergence and the discontinuity such as shock waves,
the functional can be approximated
to the function of a finite number of derivatives at the current position $x$
by substituting $X(x') \simeq X(x) + \sum_{l=1}^k (1/l!)[(x'^\mu-x^\mu)\cdot\partial_\mu]^l X(x)$:
\begin{align}
  \Gamma(x) &= \Gamma(X(x), \partial_\mu X(x), \dots, \partial_{\mu_1}\cdots\partial_{\mu_k} X(x)).
  \vv\label{eq:hydro.dissipative-deriv}
\end{align}
The {\it differential form} of the constitutive equation can be obtained
by the gradient expansion of the right-hand side of Eq.~\eqref{eq:hydro.dissipative-deriv}
with respect to the derivatives.
In hydrodynamics, this form of the constitutive equation is usually used to solve the dynamics.
It should be noted here that $\Gamma$ itself is counted as the order 1 in the gradient expansion
because the lowest order of the dissipative currents is given by the first-order terms
from the symmetry consideration.

In the Navier--Stokes theory,
using the second law of thermodynamics,
the lowest order of the dissipative currents is constrained as
\begin{align}
  \Gamma = \kappa F, \vv\label{eq:hydro.navier-stokes}
\end{align}
where the Onsager coefficients, $\kappa$, and the first-order thermodynamic force, $F$, are defined as
\begin{align}
  \kappa &\eqdef \diag(\zeta, 2\eta\Delta^{\mu\nu\alpha\beta}, -\kappa_{ij}\Delta^{\mu\alpha}), \\
  F &\eqdef \begin{pmatrix}
    -\theta \\
    \partial_{\langle\alpha}u_{\beta\rangle} \\
    -T\nabla_\alpha\frac{\mu_i}{T}
  \end{pmatrix},
\end{align}
respectively.
Here $\theta \eqdef \partial_\mu u^\mu$ and
$\circ^{\langle\mu\nu\rangle} \eqdef \Delta^{\mu\nu}{}_{\alpha\beta} \circ^{\alpha\beta}$
with $\Delta^{\mu\nu\alpha\beta}
\eqdef \frac12(\Delta^{\mu\alpha}\Delta^{\nu\beta} + \Delta^{\mu\beta}\Delta^{\nu\alpha})
- \frac13\Delta^{\mu\nu}\Delta^{\alpha\beta}$
being the projector for tensors onto the spatial symmetric traceless tensor components in the local rest frame.
The symbol $\nabla^\mu \eqdef \Delta^{\mu\nu}\partial_\nu$
denotes the spatial derivative in the local rest frame.
The Onsager coefficients $\kappa$ are thermodynamic quantities, \thatis, functions of $(e, u^\mu, n_i)$.
The Onsager coefficient matrix $\kappa$ is positive semi-definite due to the second law of thermodynamics
and also symmetric due to the Onsager reciprocal relations\footnote{
  We do not consider time-reversal odd coefficients in this paper.}.
It should noted that the factor $-\Delta^{\mu\nu}$ appearing in the diffusion component of $\kappa$
is positive semi-definite as $-\Delta^{\mu\nu} = \diag(0,1,1,1)_{\mathrm{LRF}}$ in the local rest frame.

In relativistic hydrodynamics,
the Navier--Stokes theory is known to have problems of acausal modes and numerical instabilities.
To circumvent the problems, the minimal extension to the constitutive equation would be
the simplified Israel--Stewart theory~\cite{Israel:1979wp}
which includes the term of the relaxation time as a second-order term:
\begin{align}
  \Gamma = \kappa F - \tau_R \mathcal{D} \Gamma,
  \vv\label{eq:hydro.simplified-IS}
\end{align}
where $\mathcal{D}$ denotes the substantial time derivative of the dissipative currents,
and the transport coefficient matrix, $\tau_R$, is the relaxation time.
They have the forms,
\begin{align}
  \mathcal{D} &= \Delta\D,
  \vv\label{eq:hydro.relaxation-time-D} \\
  \Delta &= \diag(1, \Delta^{\mu\nu}{}_{\alpha\beta}, \Delta^{\mu}{}_{\alpha}),
  \vv\label{eq:hydro.generic-dissipative-projector} \\
  \tau_R &= \diag(\tau_\Pi, \tau_\pi, \tau_{ij}),
  \vv\label{eq:hydro.relaxation-time-matrix}
\end{align}
where $\D \eqdef u^\mu \partial_\mu$ is the time derivative in the local rest frame,
and $\tau_\Pi$, $\tau_\pi$ and $\tau_{ij}$ are the relaxation times for
the bulk pressure, the shear stress and the diffusion currents, respectively.
The symbol $\Delta$ denotes the projectors onto the space of the dissipative current $\Gamma$,
and the symbol $\tau_R$ denotes the matrix of the relaxation time.
To make the maximal signal propagation speed smaller than the speed of light,
the relaxation times should be chosen appropriately.
Now we have the time derivative of the dissipative current itself on the right-hand side,
so the constitutive equation is regarded as a dynamical equation.
Also, one could include other second-order terms.
This kind of second-order dissipative hydrodynamics
which respects the causality
is called causal dissipative hydrodynamics.

\subsection{Integral form of constitutive equation}
\label{sec:hydro.integral-form}
To discuss the nature of the hydrodynamic fluctuations later in Sec.~\ref{sec:fdr},
we here introduce another representation of the constitutive equation, \thatis,
the {\it integral form} of the constitutive equation
which is for example obtained in the linear-response regime.
It should be noted that this particular form of the constitutive equation
is not usually used to solve the actual dynamics
because of its analytical or numerical complexity.
In this paper we consider the integral form obtained in the linear-response theory
with the following structure:
\begin{align}
  \Gamma(x) &= \int\d^4x' G(x, x') \kappa(x') F(x'),
  \vv\label{eq:hydro.integral-form}
\end{align}
where the retarded kernel, $G(x, x')\kappa(x')$, is called the memory function.
Here $\kappa$ was factored out for simplicity in the later discussion.
In this paper, $G(x, x')$ is also called the memory function.
Here physically required properties of the memory function $G(x, x')$ are summarized:
\begin{itemize}
\item The memory function should be retarded:
  \begin{align}
    G(x, x') &= 0, \quad \text{if} \quad x^0 < x'^0.
    \vv\label{eq:hydro.memory-function.retardation}
  \end{align}
  This is because to solve the dynamics
  the dissipative currents should not depend on the future information.
\item The memory function should vanish for spatially separated two points to respect the causality:
  \begin{align}
    G(x, x') &= 0, \quad \text{if} \quad (x-x')^2 < 0.
    \vv\label{eq:hydro.memory-function.causality}
  \end{align}
\item The memory function should vanish in the limit of large time:
  \begin{align}
    G(x, x') &\to 0, \quad \text{as} \quad x^0 - x'^0 \to \infty.
    \vv\label{eq:hydro.memory-function.relaxation}
  \end{align}
  The memory function should not depend on the thermodynamic forces of infinite past
  because the system near the equilibrium should forget
  the information of the fluctuations from the local equilibrium in a finite time.
\item If the background fields are homogeneous and static,
  the memory function has a translational symmetry, \thatis,
  $G(x, x') = G(x-x')$.
\end{itemize}

Some class of the differential forms of the constitutive equation
has corresponding equivalent integral forms.
We here explicitly show the integral forms
of the Navier--Stokes theory and the simplified Israel--Stewart theory
by giving the expression for the memory functions.
More general consideration on the class of the differential forms
that has corresponding integral forms will be discussed in Sec.~\ref{sec:general-differential-form}.
For the simplest example,
in the Navier--Stokes theory~\eqref{eq:hydro.navier-stokes},
the memory function is identified to be
the delta function,
\begin{align}
  G(x, x') = \delta^{(4)}(x-x').
  \vv\label{eq:hydro.memory-function-NS}
\end{align}

For another example,
the differential form~\eqref{eq:hydro.simplified-IS}
of the simplified Israel--Stewart theory
can be symbolically solved
to obtain an explicit formula for the dissipative current as
\begin{align}
  \Gamma(\tau) &= \int^\tau_{-\infty} \d\tau_1 G(\tau, \tau_1)\kappa(\tau_1) F(\tau_1),
  \vv\label{eq:hydro.integral-form-IS} \\
  G(\tau, \tau_1) &\eqsym \mathrm{T}\exp\left(-\int^{\tau}_{\tau_1}\d\tau_2 \tau_R^{-1}(\tau_2)\right) \tau_R^{-1}(\tau_1),
  \vv\label{eq:hydro.integral-form-kernel}
\end{align}
where $\tau$ is the proper time,
and $\mathrm{T}\exp$ is the $\tau$-ordered exponential for the matrix $\tau_R^{-1}(\tau_2)$.
Here we used the fact that the derivative $\D$ corresponds
to the derivative with respect to the proper time $\tau$.
In this section the symbol $\eqsym$ is used to express that the equality
gives a symbolical solution rather than the exact solution.
In solving the equation, we ignored the projectors appearing in $\mathcal{D}$ in Eq.~\eqref{eq:hydro.relaxation-time-D},
which have not been attacked directly in inhomogeneous and non-static backgrounds.
For example, to avoid the complexity in inhomogeneous and non-static backgrounds,
Ref.~\cite{Koide:2006ef} introduced in the differential form
a new term cancelling with the complicated effects of the projector,
which is not present in the Israel--Stewart theory.
In this paper, instead of modifying
the differential form of the constitutive equations
to match with the simple integral form,
we will obtain the integral form
corresponding to the Israel--Stewart theory in the linear-response regime,
in Sec.~\ref{sec:pathline}.

Finally we reconsider Eqs.~\eqref{eq:hydro.integral-form-IS}~and~\eqref{eq:hydro.integral-form-kernel} more carefully.
In particular we give explicit definitions of $\tau$
and the integration with respect to $\tau$.
The time $\tau$ is defined to be the proper time of the fluid particle,
which is a virtual particle that moves with the flow velocity.
The proper time satisfies $\D\tau = 1$, so solving the equation the proper time can be defined as
\begin{align}
  \tau(t,\bm{x})
    &\eqdef \tau(t_0, \bm{x}) + \int_{t_0}^{t} \frac{dt'}{u^0} [1-u^i\partial_i \tau(t',\bm{x})],
\end{align}
where the summation over spatial indices are taken for the upper and lower index pair of $i$.
Here the initial proper time distribution, $\tau(t_0, \bm{x})$, can be freely chosen
because the change of the initial proper time is just a time reparametrization.
Fluid particles are specified with comoving spatial coordinates $\bm{\sigma}$,
which is a counter part of the proper time $\tau$.
To satisfy $\D\bm{\sigma}=0$
so that $\bm{\sigma}$ for the fluid particle does not change in time,
$\bm{\sigma}$ is defined as
\begin{align}
  \bm{\sigma}(t,\bm{x})
    &\eqdef \bm{\sigma}(t_0, \bm{x}) - \int_{t_0}^{t} \frac{dt'}{u^0} u^i\partial_i \bm{\sigma}(t',\bm{x}).
\end{align}
The initial coordinates $\bm{\sigma}(t_0, \bm{x})$ also have the freedom degrees of choice
but should be chosen so that the spatial coordinates $\bm{x}$
and the coordinates $\bm{\sigma}$ are in one-to-one correspondence.
In this way the spacetime point $x^\mu$
and the comoving coordinates $\sigma^\mu \eqdef (\tau, \bm{\sigma})$
becomes in one-to-one correspondence as a whole,
so the comoving coordinates can be regarded
as another coordinate system to specify spacetime points.
It should be noted here that the time derivative $\D$ is reduced to
the simple partial differential with respect to $\tau$: $\D = \partial/\partial\tau$.
The trajectory or the world line of a fluid particle, which is called the {\it pathline},
can be defined by a fixed $\bm{\sigma} \equiv \const$.
The integration in Eq.~\eqref{eq:hydro.integral-form-IS} is actually
defined as the integration on a pathline:
\begin{align}
  \Gamma(\tau, \bm\sigma)
    &= \int^\tau_{-\infty} \d\tau_1 G(\tau, \tau_1; \bm{\sigma})
    \kappa(\tau_1, \bm{\sigma}) F(\tau_1, \bm{\sigma}), \\
  G(\tau, \tau_1; \bm{\sigma})
    &\eqsym \mathrm{T}\exp\left(-\int^{\tau}_{\tau_1}\d\tau_2
    \tau_R^{-1}(\tau_2, \bm{\sigma})\right) \tau_R^{-1}(\tau_1, \bm{\sigma}). 
\end{align}
The above expression of the memory function can be recasted into a compatible form
with Eq.~\eqref{eq:hydro.integral-form} as follows:
\begin{align}
  G(x, x') &\eqsym \mathrm{T}\exp\left(-\int_{\tau(x')}^{\tau(x)}\d\tau_2 \tau_R^{-1}(\tau_2, \bm{\sigma}(x))\right)
    \tau_R^{-1}(x') \theta^{(4)}(\sigma(x),\sigma(x')), \\
  \theta^{(4)}(\sigma, \sigma')
    &\eqdef \Theta(\tau - \tau') \delta^{(3)}(\bm{\sigma} - \bm{\sigma}')
    \left|\frac{\partial \sigma'^\mu}{\partial x'^\alpha}\right|,
\end{align}
where $\frac{\partial \sigma'^\mu}{\partial x'^\alpha}$ denotes the Jacobian,
and $\Theta(\tau)$ denotes the Heaviside function.

\section{Tensor structure of memory function}
\label{sec:pathline}

In Sec.~\ref{sec:hydro.integral-form}
we ignored the projectors appearing in the time derivative $\mathcal{D}$
to obtain Eq.~\eqref{eq:hydro.integral-form-kernel} for simplicity.
However, to obtain the correct memory function
we need to consider the tensor structure of the memory function
emerged from the projectors in $\mathcal{D}$.
The projectors in $\mathcal{D}$ are related to the {\it transversality}
of the dissipative currents to the flow velocity, \thatis, $u_\mu\pi^{\mu\nu}=u_\mu\nu_i^\mu = 0$,
which follows from the definition of the dissipative currents by tensor decomposition.
If we do not consider the projectors in $\mathcal{D}$
in solving Eq.~\eqref{eq:hydro.simplified-IS},
the resulting dissipative currents would break the transversality.
For example, if we do not consider the projectors in $\mathcal{D}$,
the dynamical equation~\eqref{eq:hydro.simplified-IS} for the shear-stress component would become
\begin{align}
  \tau_\pi\D\pi^{\mu\nu} = -\pi^{\mu\nu} + 2\eta\partial^{\langle\mu} u^{\nu\rangle}.
\end{align}
This constitutive equation would result in
\begin{align}
  \D(u_\mu \pi^{\mu\nu}) &= \pi^{\mu\nu}\D u_\mu
    + u_\mu \tau_\pi^{-1} (2\eta\partial^{\langle\mu} u^{\nu\rangle} - \pi^{\mu\nu})
    = \pi^{\mu\nu}\D u_\nu.
\end{align}
The right-hand side does not vanish
if the velocity fields have time dependence $\D u_\nu\neq 0$,
which means the transversality $u_\mu\pi^{\mu\nu}=0$ would be broken by the time evolution
even if the shear stress initially satisfies the transversality.
If the projectors are correctly considered, the dynamical equation becomes
\begin{align}
  \tau_\pi\Delta^{\mu\nu}{}_{\alpha\beta}\D\pi^{\alpha\beta}
  &= -\pi^{\mu\nu} + 2\eta\partial^{\langle\mu} u^{\nu\rangle},
\end{align}
and the transversality $u_\mu\pi^{\mu\nu}=0$ is preserved
under this correct version of the constitutive equation.
In fact the memory function~\eqref{eq:hydro.integral-form-kernel},
which is obtained without considering the projector,
explicitly breaks the transversality of the non-vanishing shear stress
in inhomogeneous and non-static background.
In this section we obtain the correct memory function
which preserves the transversality and also provides the solution
to the original constitutive equation~\eqref{eq:hydro.simplified-IS}
with the projectors in $\mathcal{D}$ considered.
The well-known properties of the projectors
are listed in Appendix~\ref{app:pathline.projector-property}
which will be referenced in Sec.~\ref{sec:pathline.defs}.

\subsection{Pathline projectors and their properties}
\label{sec:pathline.defs}
\newcommand{\tauI}{\tau_\mathrm{i}}
\newcommand{\tauF}{\tau_\mathrm{f}}
\newcommand{\Df}{\D_\mathrm{f}}

In this section we introduce new tensors which we call the {\it pathline projectors} in this paper.
First we give a definition of the pathline projectors and then discuss their properties.
In this section all the projectors and flow velocities
are evaluated on a single fixed pathline specified by $\bm{\sigma} = \bm{\sigma}^*$,
so we will omit the explicit dependence on $\bm{\sigma}^*$,
\thatis, $\Delta(\tau)\ud{\mu}{\alpha} \eqdef \Delta(\tau, \bm\sigma^*)\ud{\mu}{\alpha}$,
$\Delta(\tau)\ud{\mu\nu}{\alpha\beta} \eqdef \Delta(\tau, \bm\sigma^*)\ud{\mu\nu}{\alpha\beta}$,
and $u(\tau)^\mu \eqdef u(\tau, \bm{\sigma}^*)^\mu$.

First new projectors are introduced as follows:
\begin{align}
  \Delta(\tauF; \tauI)\ud{\mu}{\alpha}
    &\eqdef \lim_{N\to\infty} \Delta(\tauF)\ud{\mu}{\alpha_0}
    \biggl[\prod_{k=0}^{N-1}\Delta(\tauF + \tfrac{\tauI-\tauF}N k)\ud{\alpha_k}{\alpha_{k+1}}\biggr]
    \Delta(\tauI)\ud{\alpha_N}{\alpha},
    \vv\label{eq:pathline.def3} \\
  \Delta(\tauF; \tauI)\ud{\mu\nu}{\alpha\beta}
    &\eqdef \lim_{N\to\infty} \Delta(\tauF)\ud{\mu\nu}{\alpha_0\beta_0}
    \biggl[\prod_{k=0}^{N-1}\Delta(\tauF + \tfrac{\tauI-\tauF}N k)\ud{\alpha_k\beta_k}{\alpha_{k+1}\beta_{k+1}}\biggr]
    \Delta(\tauI)\ud{\alpha_N\beta_N}{\alpha\beta}.
    \vv\label{eq:pathline.def5}
\end{align}
With these newly introduced projectors, projections are performed at every moment on the pathline.
It should be noted that the projected spaces by
$\Delta(\tau)\ud{\mu\nu}{\alpha\beta}$ and $\Delta(\tau)\ud{\mu}{\alpha}$
are time dependent because the flow velocity $u(\tau)^\mu$
which defines the local rest frame depends on the time.

When $u^\mu$, $\D u^\mu$ and $\D^2 u^\mu$ are bounded and continuous in a considered domain,
the limits in the definitions of the pathline projectors~\eqref{eq:pathline.def3}~and~\eqref{eq:pathline.def5} are convergent,
and the projectors become well-defined.
Then the pathline projectors have the following properties:
\begin{align}
  \Delta(\tauF;\tauI)\ud\mu\alpha
    &= \Delta(\tauF;\tauI)\ud\mu\kappa \Delta(\tauI)\ud\kappa\alpha
    = \Delta(\tauF)\ud\mu\kappa \Delta(\tauF;\tauI)\ud\kappa\alpha, \vv\label{eq:ttproj3.prop-repr}\\
  \Delta(\tauF;\tauI)\ud{\mu\nu}{\alpha\beta}
    &= \Delta(\tauF;\tauI)\ud{\mu\nu}{\kappa\lambda} \Delta(\tauI)\ud{\kappa\lambda}{\alpha\beta}
    = \Delta(\tauF)\ud{\mu\nu}{\kappa\lambda} \Delta(\tauF;\tauI)\ud{\kappa\lambda}{\alpha\beta}, \vv\label{eq:ttproj5.prop-repr}\\
  \Delta(\tauI;\tauI)\ud\mu\alpha
    &= \Delta(\tauI)\ud\mu\alpha, \vv\label{eq:ttproj3.prop-same}\\
  \Delta(\tauI;\tauI)\ud{\mu\nu}{\alpha\beta}
    &= \Delta(\tauI)\ud{\mu\nu}{\alpha\beta}, \vv\label{eq:ttproj5.prop-same}\\
  \Delta(\tauF;\tauI)\ud\mu\alpha
    &= \Delta(\tauI;\tauF)\du\alpha\mu, \vv\label{eq:ttproj3.prop-sym} \\
  \Delta(\tauF;\tauI)\ud{\mu\nu}{\alpha\beta}
    &= \Delta(\tauI;\tauF)\du{\alpha\beta}{\mu\nu}, \vv\label{eq:ttproj5.prop-sym} \\
  \Delta(\tauF;\tau')\ud\mu\kappa \Delta(\tau';\tauI)\ud\kappa\alpha
    &= \Delta(\tauF;\tauI)\ud\mu\alpha, \vv\label{eq:ttproj3.prop-connect} \\
  \Delta(\tauF;\tau')\ud{\mu\nu}{\kappa\lambda} \Delta(\tau';\tauI)\ud{\kappa\lambda}{\alpha\beta}
    &= \Delta(\tauF;\tauI)\ud{\mu\nu}{\alpha\beta}, \vv\label{eq:ttproj5.prop-connect} \\
  \Df\Delta(\tauF;\tauI)\ud\mu\alpha
    &= [\Df\Delta(\tauF)\ud\mu\kappa] \Delta(\tauF;\tauI)\ud\kappa\alpha, \vv\label{eq:ttproj3.prop-dyn} \\
  \Df\Delta(\tauF;\tauI)\ud{\mu\nu}{\alpha\beta}
    &= [\Df\Delta(\tauF)\ud{\mu\nu}{\kappa\lambda}] \Delta(\tauF;\tauI)\ud{\kappa\lambda}{\alpha\beta}, \vv\label{eq:ttproj5.prop-dyn}\\
    &= [\Df\Delta(\tauF)\ud\mu\kappa g\ud\nu\lambda + g\ud\mu\kappa \Df\Delta(\tauF)\ud\nu\lambda]
      \Delta(\tauF;\tauI)\ud{\kappa\lambda}{\alpha\beta} \vv\label{eq:ttproj5.prop-dyn2},
\end{align}
where $\Df \eqdef \partial/\partial\tauF$.
For the details of the proof of the convergence of the pathline projectors
and the above properties, see Appendix~\ref{sec:pathline.proof}.
In the definition of the pathline projectors,
projections are performed at discrete times
and then the number of projections $N$ are taken to be infinity.
It is non trivial that such a limit exists
and even the time derivative can be safely applied to the limit value.
We here briefly give a sketch of the proof in Appendix~\ref{sec:pathline.proof}:
First a function sequence $P_N(\tauF;\tauI)$ indexed by $N$,
with which a pathline projector can be written as $\lim_{N\to\infty} P_N(\tauF;\tauI)$,
is introduced. Next $P_N(\tauF;\tauI)$ is shown to be a Cauchy sequence
using the fact that the time derivative of $P_N(\tauF;\tauI)$ can be
written by a regular part and the residual part scaling as $1/N$.
Then we can define the pathline projector $P(\tauF;\tauI)$ as the limit value
and show several properties which do not include the time derivatives.
Using the compact convergence of $P_N(\tauF;\tauI)$,
the time derivative $\Df P(\tauF;\tauI)$ is also obtained and shown to be convergent.

Eqs.~\eqref{eq:ttproj3.prop-repr}~and~\eqref{eq:ttproj5.prop-repr} tell that
the left (right) indices of the pathline projectors behave as
spatial vector or spatial symmetric traceless tensor at the time $\tauF$ ($\tauI$).
As consequences the following properties follow:
\begin{align}
  \Delta(\tauF;\tauI)\ud\mu\alpha u(\tauI)^\alpha &= 0,\\
  u(\tauF)_\mu \Delta(\tauF;\tauI)\ud\mu\alpha &= 0, \\
  \Delta(\tauF;\tauI)\ud{\nu\mu}{\alpha\beta}
  = \Delta(\tauF;\tauI)\ud{\mu\nu}{\beta\alpha}
    &= \Delta(\tauF;\tauI)\ud{\mu\nu}{\alpha\beta}, \\
  \Delta(\tauF;\tauI)\ud{\mu\nu}{\alpha\beta} u(\tauI)^{\alpha} &= 0,\\
  u(\tauF)_{\mu} \Delta(\tauF;\tauI)\ud{\mu\nu}{\alpha\beta} &= 0, \\
  \Delta(\tauF;\tauI)^{\mu\nu}{}_\alpha{}^\alpha =
  \Delta(\tauF;\tauI)\ud{\mu\nu}{\alpha\beta} \Delta(\tauI)^{\alpha\beta} &= 0,\\
  \Delta(\tauF;\tauI)^\mu{}_{\mu\alpha\beta} =
  \Delta(\tauF)_{\mu\nu} \Delta(\tauF;\tauI)\ud{\mu\nu}{\alpha\beta} &= 0.
\end{align}

Using Eqs.~\eqref{eq:ttproj3.prop-same}--\eqref{eq:ttproj5.prop-connect},
it can be shown that
a pathline projection preserves
the norms of a spatial vector and a spatial symmetric traceless tensor:
\begin{align}
  (\Delta(\tauF;\tauI)\ud\mu\alpha A^\alpha)
  (\Delta(\tauF;\tauI)_\mu{}^\beta A_\beta)
    &= A^\alpha A_\alpha,
    \vv\label{eq:ttproj3.prop-preserve-norm} \\
  (\Delta(\tauF;\tauI)\ud{\mu\nu}{\alpha\beta} B^{\alpha\beta})
  (\Delta(\tauF;\tauI)_{\mu\nu}{}^{\gamma\delta} B_{\gamma\delta})
    &= B^{\alpha\beta} B_{\alpha\beta}
    \vv\label{eq:ttproj5.prop-preserve-norm}
\end{align}
for any spatial vector $A^\mu$
such that $\Delta(\tauI)\ud\mu\alpha A^\alpha = A^\mu$
and any spatial symmetric traceless tensor $B^{\mu\nu}$
such that $\Delta(\tauI)\ud{\mu\nu}{\alpha\beta} B^{\alpha\beta} = B^{\mu\nu}$.
This means that a pathline projection serves as a certain Lorentz transformation
for spatial vectors and spatial symmetric traceless tensors.

By solving Eqs.\eqref{eq:ttproj3.prop-dyn}~and~\eqref{eq:ttproj5.prop-dyn}
with the initial conditions~\eqref{eq:ttproj3.prop-same}~and~\eqref{eq:ttproj5.prop-same},
another representation of the projectors can be obtained as
\begin{align}
  \Delta(\tauF;\tauI)\ud\mu\alpha
    &= \left[{\rm T}\exp\left(\int_{\tauI}^{\tauF} d\tau \D\Delta(\tau)\ud\mu\kappa \right)\right]^\mu_{\hphantom{\mu}\kappa}
      \Delta(\tauI)\ud\kappa\alpha, \\
  \Delta(\tauF;\tauI)\ud{\mu\nu}{\alpha\beta}
    &= \left[{\rm T}\exp\left(\int_{\tauI}^{\tauF} d\tau \D\Delta(\tau)\ud{\mu\nu}{\kappa\lambda} \right)\right]^{\mu\nu}_{\hphantom{\mu\nu}\kappa\lambda}
      \Delta(\tauI)\ud{\kappa\lambda}{\alpha\beta},
\end{align}
where $\mathrm{T}\exp\int d\tau \circ$ is
a $\tau$-ordered exponential,
\thatis, the contractions of the Lorentz indices of
$\Delta(\tau)\ud\mu\lambda$ and $\Delta(\tau)\ud{\mu\nu}{\kappa\lambda}$
are performed after the time ordering.

It is known from Eqs.~\eqref{eq:ttproj5.prop-dyn2} that
the spatial traceless symmetric pathline projector, $\Delta(\tauF;\tauI)\ud{\mu\nu}{\alpha\beta}$,
can be simply written in terms of
the spatial pathline projector, $\Delta(\tauF;\tauI)\ud{\mu}{\alpha}$, as
\begin{align}
  \Delta(\tauF;\tauI)\ud{\mu\nu}{\alpha\beta}
  &= \Delta(\tauF;\tauI)\ud\mu\kappa
    \Delta(\tauF;\tauI)\ud\nu\lambda
    \Delta(\tauI)\ud{\kappa\lambda}{\alpha\beta}.
\end{align}
This reflects the fact that
the tracelessness and the symmetry of the tensor
are independent of the local rest frame $u^\mu$.
Once the projection into the traceless symmetric components is made,
there is no need of such projection in other times.

\subsection{Memory function with tensor structures}
\label{sec:pathline.memory-function}

Now we are ready to give a solution to
the second-order constitutive equation~\eqref{eq:hydro.simplified-IS} using pathline projectors
which replaces the symbolic solution~\eqref{eq:hydro.integral-form-kernel} breaking the transversality.
Using the properties~\eqref{eq:ttproj3.prop-dyn}--\eqref{eq:ttproj5.prop-dyn2},
the following equations can be shown:
\begin{gather}
  \Delta(\tau)\ud\mu\kappa
    \D \Delta(\tau;\tau_0)\ud\kappa\alpha =0, \\
  \Delta(\tau)\ud{\mu\nu}{\kappa\lambda}
    \D \Delta(\tau;\tau_0)\ud{\kappa\lambda}{\alpha\beta}
  = \Delta(\tau)\ud\mu\kappa
    \Delta(\tau)\ud\nu\lambda
    \D \Delta(\tau;\tau_0)\ud{\kappa\lambda}{\alpha\beta} =0.
\end{gather}
These equations can be used to solve a certain class  
of the differential form of the constitutive equations.

For example,
if a spatial vector $A^\mu(\tau)$ and a spatial traceless symmetric tensor $B^{\mu\nu}(\tau)$
obey the dynamical equations:
\begin{align}
  \Delta\ud{\mu}{\kappa}
    \D A^\kappa &=0, \\
  \Delta\ud{\mu\nu}{\kappa\lambda}
    \D B^{\kappa\lambda} &=0,
\end{align}
the solutions are written as
\begin{align}
  A^\mu(\tau)
    &= \Delta(\tau;\tau_0)\ud{\mu}{\kappa}A^\kappa(\tau_0),
    \vv\label{eq:ttproj3.simple-equation-solution} \\
  B^{\mu\nu}(\tau)
    &= \Delta(\tau;\tau_0)\ud{\mu\nu}{\kappa\lambda} B^{\kappa\lambda}(\tau_0),
    \vv\label{eq:ttproj5.simple-equation-solution}
\end{align}
with $A^\kappa(\tau_0)$ and $B^{\kappa\lambda}(\tau_0)$ being initial conditions.

For another example, let us consider the second-order constitutive equations of the form:
\begin{align}
  \pi^{\mu\nu}
  + \tau_\pi\Delta\ud{\mu\nu}{\kappa\lambda}
    \D\pi^{\kappa\lambda} &= X^{\mu\nu},
    \vv\label{eq:pathline.memfun.ce-X.shear} \\
  \nu_i^\mu
  + \sum_{j=1}^n \tau_{ij} \Delta\ud{\mu}{\kappa}
    \D \nu_j^\kappa &= X_i^{\mu},
    \vv\label{eq:pathline.memfun.ce-X.diffuse}
\end{align}
with $X^{\mu\nu}$ and $X_i^\mu$ being
miscellaneous second-order terms
which do not contain dissipative currents.
The equations can be solved with respect to the dissipative currents
using the pathline projectors as
\begin{align}
  \pi^{\mu\nu}(\tau)
    &= G\ud{\mu\nu}{\alpha\beta}(\tau,\tau_0)\tau_\pi(\tau_0)\pi^{\alpha\beta}(\tau_0)
      + \int_{\tau_0}^\tau \d\tau_1 G\ud{\mu\nu}{\alpha\beta}(\tau,\tau_1) X^{\alpha\beta}(\tau_1),
    \vv\label{eq:pathline.memfun.ce-X-solution.shear} \\
  \nu_i^\mu(\tau)
    &= \sum_{j,k=1}^n
      G_{ij}{}\ud\mu\alpha(\tau,\tau_0) \tau_{jk}(\tau_0) \nu_k^\alpha(\tau_0)
      + \sum_{j=1}^n \int_{\tau_0}^\tau \d\tau_1
      G_{ij}{}\ud\mu\alpha(\tau,\tau_1) X_j^\alpha(\tau_1),
    \vv\label{eq:pathline.memfun.ce-X-solution.diffuse} \\
  G\ud{\mu\nu}{\alpha\beta}(\tau,\tau_1)
    &= \exp\left(-\int_{\tau_1}^\tau\frac{\d\tau_2}{\tau_\pi(\tau_2)}\right) \frac1{\tau_\pi(\tau_1)}
      \Delta(\tau;\tau_1)\ud{\mu\nu}{\alpha\beta},
    \vv\label{eq:pathline.memfun.ce-X-mem.shear} \\
  G_{ik}{}\ud\mu\alpha(\tau,\tau_1)
    &= \sum_{j=1}^n
      \left[\T\exp\left(-\int_{\tau_0}^\tau \d\tau_2 \tau^{-1}_{ij}(\tau_2)\right)\right]_{ij}
      \tau_{jk}^{-1}(\tau_1) \Delta(\tau;\tau_1)\ud\mu\alpha,
    \vv\label{eq:pathline.memfun.ce-X-mem.diffuse}
\end{align}
where $\pi^{\alpha\beta}(\tau_0)$ and $\nu_i^\alpha(\tau_0)$ are initial conditions.

\subsection{Memory function in simplified Israel--Stewart theory}
\label{sec:pathline.simplified-IS}

To find the memory functions for the shear stress tensor and diffusion currents
in the simplified Israel--Stewart theory~\eqref{eq:hydro.simplified-IS},
we can consider the case $X^{\mu\nu} = 2\eta\partial^{\langle\mu} u^{\mu\rangle}$
and $X_i^\mu = -T\nabla^\mu (\mu_i/T)$
and take the limit $\tau_0 \to -\infty$:
\begin{align}
  \pi^{\mu\nu}(\tau)
    &= \int^\tau_{-\infty} \d\tau_1 G\ud{\mu\nu}{\alpha\beta}(\tau,\tau_1) 2\eta\partial^{\langle\alpha} u^{\beta\rangle}(\tau_1), \\
  \nu_i^\mu(\tau)
    &= -\sum_{j,k=1}^n \int_{-\infty}^\tau \d\tau_1 G_{ij}{}\ud\mu\alpha(\tau,\tau_1) \kappa_{jk} T(\tau_1)\nabla^\alpha \frac{\mu_k(\tau_1)}{T(\tau_1)}.
\end{align}
Thus the expression for the memory function~\eqref{eq:hydro.integral-form-kernel}
which correctly takes account of the tensor structure is
\begin{align}
  G(x,x') &= \diag(
    G_\Pi(x,x'),
    G\ud{\mu\nu}{\alpha\beta}(x,x'),
    G_{ij}{}\ud\mu\alpha(x,x')),
  \vv\label{sec:pathline.simplified-IS.memory}
\end{align}
where
\begin{align}
  G_\Pi(x,x')
    &= \exp\left(-\int_{\tau(x')}^{\tau(x)} \frac{\d\tau_2}{\tau_\Pi(\tau_2,\bm{\sigma}(x))}\right)
    \frac1{\tau_\Pi(x')} \theta^{(4)}(\sigma(x), \sigma(x')),
    \vv\label{sec:pathline.simplified-IS.memory-bulk} \\
  G\ud{\mu\nu}{\alpha\beta}(x,x')
    &= \exp\left(-\int_{\tau(x')}^{\tau(x)} \frac{\d\tau_2}{\tau_\pi(\tau_2,\bm{\sigma}(x))}\right)
      \frac1{\tau_\pi(x')}
      \Delta(\tau(x);\tau(x'),\bm{\sigma}(x))\ud{\mu\nu}{\alpha\beta} \theta^{(4)}(\sigma(x), \sigma(x')),
    \vv\label{sec:pathline.simplified-IS.memory-shear} \\
  G_{ik}{}\ud\mu\alpha(x,x')
    &= \sum_{j=1}^n
      \left[\T\exp\left(-\int_{\tau(x')}^{\tau(x)} \d\tau_2 \tau^{-1}_{ij}(\tau_2,\bm{\sigma}(x))\right)\right]_{ij}
      \tau_{jk}^{-1}(x') \Delta(\tau(x);\tau(x'),\bm{\sigma}(x))\ud\mu\alpha \theta^{(4)}(\sigma(x), \sigma(x')).
    \vv\label{sec:pathline.simplified-IS.memory-diffusion}
\end{align}
By defining a pathline projector for the generic dissipative currents as
\begin{align}
  \Delta(\tauF;\tauI, \bm{\sigma}) &= \diag(1,\ 
    \Delta(\tauF;\tauI,\bm{\sigma})\ud{\mu\nu}{\alpha\beta},\ 
    \Delta(\tauF;\tauI,\bm{\sigma})\ud{\mu}{\alpha}),
\end{align}
these memory functions can be summarized in the following form:
\begin{align}
  G(x,x')
    &= \left[\T\exp\left(-\int_{\tau(x')}^{\tau(x)} \d\tau_2 \tau^{-1}_R(\tau_2,\bm{\sigma}(x))\right)\right]
      \tau_R^{-1}(x') \Delta(\tau(x);\tau(x'),\bm{\sigma}(x)) \theta^{(4)}(\sigma(x), \sigma(x')).
\end{align}

Finally let us here give an expression of the memory function
in the global equilibrium for later usage.
In homogeneous and static backgrounds where $u^\mu = (1,0,0,0)$ and $T = \mu_i = \const$,
the expression of the memory function simplifies as
\begin{align}
  G(x-x') &= e^{-(x^0-x'^0)\tau^{-1}_R}
      \tau_R^{-1} \Delta \Theta(x^0-x'^0) \delta^{(3)}(\bm{x} - \bm{x}').
  \vv\label{eq:pathline.simplified-IS.generic-memory}
\end{align}
Components of the above memory function can be written down as
\begin{align}
  G_\Pi(x-x')
    &= e^{-\frac{x^0-x'^0}{\tau_\Pi}}\frac1{\tau_\Pi} \Theta(x^0 - x'^0) \delta^{(3)}(\bm{x} - \bm{x}'),
  \vv\label{eq:pathline.simplified-IS.bulk-memory} \\
  G\ud{\mu\nu}{\alpha\beta}(x-x')
    &= e^{-\frac{x^0-x'^0}{\tau_\pi}}\frac1{\tau_\pi}
      \Delta\ud{\mu\nu}{\alpha\beta} \Theta(x^0 - x'^0) \delta^{(3)}(\bm{x} - \bm{x}'),
  \vv\label{eq:pathline.simplified-IS.shear-memory} \\
  G_{ik}{}\ud\mu\alpha(x-x')
    &= \sum_{j=1}^n
      \left[e^{-(x^0-x'^0)\tau^{-1}_{ij}}\right]_{ij}
      \tau_{jk}^{-1} \Delta\ud\mu\alpha \Theta(x^0-x'^0) \delta^{(3)}(\bm{x} - \bm{x}').
  \vv\label{eq:pathline.simplified-IS.diffusion-memory}
\end{align}

%%%%%%%%%%%%%%%%%%%%%%%%%%%%%%%%%%%%%%%%%%%%%%%%%%%%%%%%%%%%%%%%%%%%%%%%%%%%%%%
\section{Causal hydrodynamic fluctuations in equilibrium}
\vv\label{sec:fdr}

\subsection{Colored noise in integral form}
\label{sec:fdr.integral-form}

So far we have considered the constitutive equations which determine
the dissipative currents in terms of the present and past hydrodynamic fields.
However such deterministic constitutive equations merely describe
the ``average'' behavior of the dissipative currents.
If the scale of interest is not sufficiently larger than the scale of the microscopic dynamics,
the deviation from the ``average'' affects the relevant dynamics.
Such deviation is nothing but the hydrodynamic fluctuations,
\thatis, the thermal fluctuations of the dissipative currents:
\begin{align}
  \delta\Gamma(x) &= \begin{pmatrix}
    \delta\Pi(x)\\
    \delta\pi^{\mu\nu}(x)\\
    \delta\nu_i^\mu(x)
  \end{pmatrix} \eqdef \Gamma(x) - \int \d^4x' G(x,x')\kappa(x')F(x').
\end{align}
The hydrodynamic fluctuations
originate from the dynamics of the microscopic degrees of freedom
and therefore cannot be uniquely determined by the macroscopic information.
Instead, the hydrodynamic fluctuations
are treated as stochastic fields whose distribution
is determined by the macroscopic information.
The dissipative currents are calculated
by the sum of the average part and the stochastic fields:
\begin{align}
  \Gamma(x) &= \int \d^4x' G(x,x')\kappa(x')F(x') + \delta\Gamma(x).
  \vv\label{eq:fdr.integral-form-with-noise}
\end{align}
Now the hydrodynamic equations are
stochastic partial differential equations (SPDE)
which are similar to the Langevin equation for the Brownian motion.
Such hydrodynamics with hydrodynamic fluctuations is called fluctuating hydrodynamics.
The major difference to the Langevin equation is that
the dynamical variable is fluid fields
instead of a position of particle,
so the noise terms become fields which have spatial dependence.

The distribution of the hydrodynamic fluctuations,
$\delta\Gamma(x)$, can be characterized by its moments
such as average values and variances.
The average values of the distribution
are zero by definition: $\langle\delta\Gamma(x)\rangle = 0$.
The variance--covariance matrix\footnote{
  Here $\langle\delta\Gamma(x)\delta\Gamma(x')^\T\rangle$
  is regarded as a ``matrix'' whose indices include spatial indices $x$ and $x'$
  as well as the usual indices that specifies the component of $\delta\Gamma$.
} of the distribution,
$\langle\delta\Gamma(x)\delta\Gamma(x')^\T\rangle$,
determines the magnitude of the noises
and is given in terms of macroscopic quantities
by the FDR based on the linear-response theory~\cite{Kubo:1957mj}.
The FDR can be obtained by the linear response theory
by assuming Onsager's regression hypothesis~\cite{Onsager:1931}:
The relaxation of spontaneous fluctuations on average behaves
as the same as the relaxation of the deviation caused by external fields
(see Ref.~\cite{MARCONI2008111} for a summary of topics and discussions on the FDR).

In this section we consider the hydrodynamic fluctuations in an equilibrium state,
\thatis, in homogeneous and static backgrounds.
In equilibrium, the FDR is given by
\begin{align}
  \langle\delta\Gamma(x)\delta\Gamma(x')^\T\rangle \Theta(x^0 - x'^0) &= G(x - x') \kappa T,
  \vv\label{eq:fdr.uniform-fdr-theta}
\end{align}
or equivalently,
\begin{align}
  \langle\delta\Gamma(x)\delta\Gamma(x')^\T\rangle
    &= T[G(x - x') \kappa + \kappa G(x' - x)^\T],
  \vv\label{eq:fdr.uniform-fdr}
\end{align}
where we used the Onsager reciprocal relation
$\kappa = \kappa^\T$ to remove the transpose of the Onsager coefficient $\kappa$.

Although we do not assume
the Gaussian distribution of the hydrodynamic fluctuations
in the following discussion,
it should be noted here that the distribution of the hydrodynamic fluctuations
is usually assumed to have the multivariate Gaussian distribution of the form,
\begin{align}
  \Pr[\delta\Gamma(x)] &\propto \exp\left(-\frac12\int\d^4x\int\d^4x' \delta\Gamma(x)^\T C^{-1}(x,x') \delta\Gamma(x')\right),
\end{align}
where $\Pr[\delta\Gamma(x)]$ is the probability density functional that a specific history of $\delta\Gamma(x)$ is realized,
and $C^{-1}(x,x')$ is a matrix-valued function and is symmetric, \thatis, $C^{-1}(x,x') = C^{-1}(x',x)$.
The reasoning of the Gaussian assumption is that,
if the system is sufficiently large so that the dissipative currents
are sum of many independent contributions from microscopic processes,
the resulting distribution becomes Gaussian due to the central limit theorem.
The freedom degrees of the multivariate Gaussian distribution,
$C^{-1}(x,x')$, are totally fixed by the autocorrelation given by the FDR through the relation:
\begin{align}
  \int \d^4x_2 \langle\delta\Gamma(x_1)\delta\Gamma(x_2)^\T\rangle C^{-1}(x_2,x_3)
  &= \delta^{(4)}(x_1 - x_3).
\end{align}
Because in this way the FDR settles
the basic statistical nature of the hydrodynamic fluctuations
from the macroscopic information,
it can be regarded as the most important relation
in considering the hydrodynamic fluctuations.

Here let us demonstrate examples of the FDR for the Navier--Stokes theory
and the simplified Israel--Stewart theory.
The FDR in the Navier--Stokes theory is obtained
by combining Eq.~\eqref{eq:hydro.memory-function-NS} and Eq.~\eqref{eq:fdr.uniform-fdr} as
\begin{align}
  \langle\delta\Gamma(x)\delta\Gamma(x')^\T\rangle
    &= 2T\kappa\delta^{(4)}(x-x').
\end{align}
From the above expression, the diagonal components read
\begin{align}
  \langle\delta\Pi(x)\delta\Pi(x')\rangle
    &= 2T\zeta \delta^{(4)}(x-x'),
    \vv\label{eq:fdr.uniform-NS.shear} \\
  \langle\delta\pi^{\mu\nu}(x)\delta\pi^{\alpha\beta}(x')\rangle
    &= 4T\eta \Delta^{\mu\nu\alpha\beta} \delta^{(4)}(x-x'),
    \vv\label{eq:fdr.uniform-NS.bulk} \\
  \langle\delta\nu_i^\mu(x)\delta\nu_j^\alpha(x')\rangle
    &= -2T\kappa_{ij} \Delta^{\mu\alpha}\delta^{(4)}(x-x').
    \vv\label{eq:fdr.uniform-NS.diffusion} \\
\end{align}
The off-diagonal components vanish:
\begin{align}
  \langle\delta\Pi(x)\pi^{\alpha\beta}(x')\rangle &=
  \langle\delta\pi^{\mu\nu}(x)\delta\nu_j^\alpha(x')\rangle =
  \langle\delta\nu_i^\mu(x)\delta\Pi(x')\rangle = 0.
\end{align}
One can observe that the autocorrelations of the noise fields are delta functions
which correspond to constant power spectra in the Fourier space.
This type of noise fields is called {\it white noise}
because modes of all the frequencies
are contained in the spectrum with an equal strength.

In the simplified Israel--Stewart theory, the FDR is obtained
from Eq.~\eqref{eq:pathline.simplified-IS.generic-memory}
and Eq.~\eqref{eq:fdr.uniform-fdr}.
\begin{align}
  \langle\delta\Gamma(x)\delta\Gamma(x')^\T\rangle
    &= T \biggl\{
      e^{-(x^0-x'^0)\tau^{-1}_R} \tau_R^{-1}\kappa\Theta(x^0-x'^0) \nonumber \\ & \qquad
      + \left[e^{-(x'^0-x^0)\tau^{-1}_R} \tau_R^{-1}\kappa\Theta(x'^0-x^0)\right]^\T
    \biggr\} \Delta \delta^{(3)}(\bm{x} - \bm{x}').
  \vv\label{eq:fdr.simplified-IS.fdr}
\end{align}
The diagonal components read
\begin{align}
  \langle\delta\Pi(x)\delta\Pi(x')\rangle
    &= 2T\zeta \frac1{\tau_\Pi} e^{-\frac{|x^0-x'^0|}{\tau_\Pi}} \delta^{(3)}(\bm{x} - \bm{x}'), \\
  \langle\delta\pi^{\mu\nu}(x)\delta\pi^{\alpha\beta}(x')\rangle
    &= 4T\eta \frac1{\tau_\pi} e^{-\frac{|x^0-x'^0|}{\tau_\pi}}
      \Delta^{\mu\nu\alpha\beta} \delta^{(3)}(\bm{x} - \bm{x}'), \\
  \langle\delta\nu_i^\mu(x)\delta\nu_j^\alpha(x')\rangle
    &= -T \sum_{k,l=1}^n
      \biggl\{ \left[e^{-(x^0-x'^0)\tau^{-1}_{ik}}\right]_{ik} \tau_{kl}^{-1} \kappa_{lj} \Theta(x^0-x'^0) \nonumber \\ &\qquad
      + \left[e^{-(x'^0-x^0)\tau^{-1}_{jk}}\right]_{jk} \tau_{kl}^{-1} \kappa_{li} \Theta(x'^0-x^0) \biggr\}
      \Delta^{\mu\alpha} \delta^{(3)}(\bm{x} - \bm{x}').
\end{align}
The off-diagonal components vanish similarly to the case of the Navier--Stokes theory.
One notices that, unlike in the Navier--Stokes theory,
the memory function is no longer the delta function.
The noise fields have finite time correlation
with the time scale of the relaxation time $\tau_R$.
This type of noise fields is called {\it colored noise} because
the noise power spectrum in the Fourier space has a characteristic frequency
corresponding to the inverse of the relaxation time.
In general, causal dissipative hydrodynamics has non-vanishing relaxation times,
so the noise fields appearing in the integral form of the constitutive equation
are always colored in causal dissipative hydrodynamics.

\subsection{White noise in differential form in simplified Israel--Stewart theory}
\label{sec:fdr.simplified-IS}

In Sec.~\ref{sec:fdr.integral-form} we introduced noise terms $\delta\Gamma(x)$
in the integral form of the constitutive equation
that gives an explicit form of the dissipative currents.
While, in the actual analysis of causal dissipative hydrodynamics,
we usually use the differential form of the constitutive equation
that gives dynamical equations for the dissipative currents.
We here consider how the noise term $\xi(x)$ enters
the differential form of the constitutive equation
and what is the nature of the fluctuations.

For a trivial example, in the case of the Navier--Stokes theory~\eqref{eq:hydro.navier-stokes},
the noise terms in the two forms are the same: $\xi(x) = \delta\Gamma(x)$
because the differential and integral forms are equivalent in this theory.
Thus the autocorrelation of $\xi(x)$ is directly given by the FDR in the integral form
\eqref{eq:fdr.uniform-NS.shear}--\eqref{eq:fdr.uniform-NS.diffusion}.

In the differential forms of higher-order theories,
the noise term, $\xi$, can be defined
as the deviation of the right-hand side of the constitutive equation from the left-hand side
similarly in the integral form.
For example, in the case of the simplified Israel-Stewart theory \eqref{eq:hydro.simplified-IS},
the noise term is defined as
\begin{align}
   \xi(x) = \begin{pmatrix}
      \xi_\Pi(x) \\ \xi_\pi^{\mu\nu}(x) \\ \xi_i^\mu(x)
    \end{pmatrix} \eqdef \Gamma(x) - [\kappa F(x) - \tau_R\mathcal{D}\Gamma(x)].
\end{align}
By substituting \eqref{eq:fdr.integral-form-with-noise}, one finds the following relation:
\begin{align}
  \xi(x) = (1 + \tau_R\mathcal{D}) \delta\Gamma(x).
\end{align}
Because $\xi(x)$ is linear with respect to $\delta\Gamma(x)$,
$\xi(x)$ is also a Gaussian noise if $\delta\Gamma(x)$ is a Gaussian noise.
The autocorrelation of $\xi(x)$ can be calculated
from the FDR of $\delta\Gamma(x)$~\eqref{eq:fdr.uniform-fdr}:
\begin{align}
  \langle\xi(x)\xi(x')^\T\rangle
  &= \langle[(1+\tau_R\mathcal{D})\delta\Gamma(x)] [(1+\tau_R\mathcal{D}')\delta\Gamma(x')]^\T\rangle \\
  &= (1+\tau_R\mathcal{D}) \langle\delta\Gamma(x) \delta\Gamma(x')^\T\rangle (1+\tau_R\overleftarrow{\mathcal{D}'})^\T \\
  &= (1+\tau_R\mathcal{D}) [G(x - x') \kappa T + T \kappa G(x'-x)^\T] (1+\tau_R\overleftarrow{\mathcal{D}'})^\T,
\end{align}
where $\mathcal{D}$ and $\mathcal{D}'$ are the time derivatives with respect to $x$ and $x'$, respectively.
The symbol $\overleftarrow{\mathcal{D}'}$ denotes that the derivative operates on the left.
Using the relation $(1+\tau_R\mathcal{D}) G(x-x') = \delta^{(4)}(x-x')$, one obtains
\begin{align}
  \langle\xi(x)\xi(x')^\T\rangle
  &= \delta^{(4)}(x - x') \kappa T (1+\tau_R\overleftarrow{\mathcal{D}'})^\T
  + (1+\tau_R\mathcal{D}) T \kappa \delta^{(4)}(x'-x) \\
  &= 2T\kappa \delta^{(4)}(x - x')
    + 2T [\tau_R\kappa]^\mathrm{A} \D \delta^{(4)}(x - x'),
  \vv\label{eq:fdr.uniform-IS.differential-fdr}
\end{align}
where $M^\mathrm{A} \eqdef (1/2)(M - M^\T)$ denotes
the antisymmetric part of the matrix $M$.
From the diagonal components one finds
\begin{align}
  \langle\xi_\Pi(x)\xi_\Pi(x')\rangle
    &= 2T\zeta \delta^{(4)}(x-x'),
  \vv\label{eq:fdr.uniform-IS.differential-fdr-shear} \\
  \langle\xi_\pi^{\mu\nu}(x)\xi_\pi^{\alpha\beta}(x')\rangle
    &= 4T\eta \Delta^{\mu\nu\alpha\beta} \delta^{(4)}(x-x'), \\
  \langle\xi_i^\mu(x)\xi_j^\alpha(x')\rangle
    &= -2T\kappa_{ij} \Delta^{\mu\alpha}\delta^{(4)}(x-x')
      -T \sum_{k=1}^n (\tau_{ik} \kappa_{kj} - \tau_{jk} \kappa_{ki}) \Delta^{\mu\alpha} \D \delta^{(4)}(x-x').
  \vv\label{eq:fdr.uniform-IS.differential-fdr-diffusion}
\end{align}
The non-diagonal components vanish so that there are no correlation between different types of noises:
$\langle\xi_\Pi(x)\xi_\pi^{\alpha\beta}(x')\rangle = \langle\xi_\pi^{\mu\nu}(x)\xi_j^\alpha(x')\rangle = \langle\xi_i^\mu(x)\xi_\Pi(x')\rangle = 0$.

One notices that the second term of Eq.~\eqref{eq:fdr.uniform-IS.differential-fdr},
which is proportional to $[\tau_R \kappa]^\mathrm{A}$, vanishes
for the shear and bulk noise, $\xi_\Pi(x)$ and $\xi_\pi^{\mu\nu}(x)$,
while it remains for the diffusion noise, $\xi_i^\mu(x)$.
In general the second term vanishes
when we consider the single component diffusion current,
or when the dissipative currents do not mix with one another due to some symmetry,
\thatis, $\tau_R$ and $\kappa$ are diagonal.
The bulk pressure is a single-component dissipative current
so that the second term trivially vanishes.
Also, in the case of the shear stress, the second term vanishes
because $\tau_R = \tau_\pi$ is scalar
and $\kappa$ is symmetric due to the Onsager's reciprocal relation
so that $[\tau_R\kappa]^{\mathrm{A}} = \tau_\pi \kappa^{\mathrm{A}}=0$.
If the second term is present,
it is difficult to interpret as the autocorrelation of $\xi(x)$
because the second term contains the derivatives on the delta function, $\D\delta^{(4)}(x-x')$.
This implies that the differential form of the constitutive equation
cannot be used to describe the hydrodynamic fluctuations
when the dissipative currents mix with one another
through the relaxation time $\tau_R$ and the Onsager coefficients $\kappa$.
In such a case, we should consider solving the integral form instead of the differential form.

When the dissipative currents do not mix with one another,
one notices that the differential form of the FDR of the simplified Israel--Stewart theory,
Eqs.~\eqref{eq:fdr.uniform-IS.differential-fdr-shear}--\eqref{eq:fdr.uniform-IS.differential-fdr-diffusion},
match with the FDR of the Navier--Stokes theory,
Eqs.~\eqref{eq:fdr.uniform-NS.shear}--\eqref{eq:fdr.uniform-NS.diffusion}.
In particular,
in spite of the fact that the noise in the integral form $\delta\Pi(x)$ is colored,
the noise in the differential form $\xi(x)$ becomes white.

%%%%%%%%%%%%%%%%%%%%%%%%%%%%%%%%%%%%%%%%%%%%%%%%%%%%%%%%%%%%%%%%%%%%%%%%%%%%%%
\subsection{General linear-response differential form}
\label{sec:general-differential-form}

In Sec.~\ref{sec:fdr.simplified-IS}
we argued that in causal dissipative hydrodynamics,
the noise in the integral form of the constitutive equations is colored
reflecting the non-vanishing relaxation time required to maintain the causality.
However, in the simplified Israel--Stewart theory~\eqref{eq:hydro.simplified-IS},
we found that the noise in the differential form is still white
despite of the colored noise in the integral form.
In addition the FDR for the differential form is exactly
the same with that of the Navier--Stokes theory~\eqref{eq:hydro.navier-stokes}.
This result appears to be non-trivial because
the contribution from the relaxation time seems
accidentally cancel with each other in Eq.~\eqref{eq:fdr.uniform-IS.differential-fdr}
when the dissipative currents do not mix with one another.
Here the question is whether
there is a higher-order theory in which
the noise of the differential form has a different form from the Navier--Stokes theory or not.
In this section, we introduce the hydrodynamic fluctuations in the general differential form in linear-response regime
and consider its FDR\@.
Next, we discuss
restrictions on the general differential form
under the condition that the dissipative currents do not mix with one another.

First the Fourier representation of the integral form
and the differential form in the linear-response regime are introduced.
In equilibrium the integral form~\eqref{eq:fdr.integral-form-with-noise}
is written by the convolution of the memory function and the thermodynamic force:
\begin{align}
  \Gamma(x) &= \int\d^4x' G(x-x') \kappa F(x') + \delta\Gamma(x).
\end{align}
In this section we take the coordinates $(t, \bm{x})$
in the local rest frame where $u^\mu = (1, 0, 0, 0)^\T$.
In the Fourier representation, the above expression is transformed to
\begin{align}
  \Gamma_{\omega,\bm{k}} &= G_{\omega,\bm{k}} \kappa F_{\omega,\bm{k}} + \delta\Gamma_{\omega,\bm{k}}.
  \vv\label{eq:fdr.integral-form.fourier}
\end{align}
Here the Fourier transform is defined as
\begin{align}
  f_{\omega,\bm{k}}
    &= \int\d^4{x} e^{ik^\mu x_\mu} f(x)
    = \int\d^4{x} e^{i\omega t - i\bm{k}\cdot\bm{x}} f(x), \\
  f(x)
    &= \int \frac{\d^4{k}}{(2\pi)^4} e^{-ik^\mu x_\mu} f_{\omega,\bm{k}}
    = \int \frac{\d^4{k}}{(2\pi)^4} e^{-i\omega t + i\bm{k}\cdot\bm{x}} f_{\omega,\bm{k}},
\end{align}
where $k^\mu = (\omega,\bm{k})$.

The differential form in the linear-response (without noise) can generally be written in the following form%
\cite{Denicol:2011fa,Denicol:2011rn}:
\begin{align}
  L_0(\D,\nabla^\mu) \Gamma(x) &=  M_0(\D,\nabla^\mu) \kappa F(x),
\end{align}
where $L_0(z,\bm{w})$ and $M_0(z,\bm{w})$ are
matrix-valued polynomials of $z$ and $\bm{w}$.
The time derivatives $\D$ in the polynomial $M_0$
are acting on the thermodynamic forces $F(x)$
which are written in terms of the thermodynamic fields $(u^\mu, e, n_i)$.
Substituting the hydrodynamic equations to ($\D u^\mu$, $\D e$, $\D n_i$) in the right-hand side,
the time derivatives $\D$ in $M_0$ can be replaced by the spatial derivatives
acting on dissipative currents or thermodynamic fields.
If non-linear terms in the hydrodynamic equations
are ignored in the linear-response regime,
the above differential form can be reorganized into the following linear form:
\begin{align}
  L(\D,\nabla^\mu) \Gamma(x) &= M(\nabla^\mu) \kappa F(x) + \xi(x),
  \vv\label{eq:fdr.higher-order.differential-form}
\end{align}
where the hydrodynamic fluctuations $\xi(x)$ are introduced here.
Here the polynomials $L(z,\bm{w})$ and $M(\bm{w})$ are required to have the following several properties:
\begin{itemize}
\item
  The polynomials $L(z,\bm{w})$ and $M(\bm{w})$ are normalized to be coprime, \thatis,
  the two polynomials do not have common factors.
  If they have common factors, they can be eliminated
  by dividing both sides of Eq.~\eqref{eq:fdr.higher-order.differential-form}.
\item
  The coefficients of the polynomials
  $L(z,\bm{w})$ and $M(\bm{w})$
  are the transport coefficients
  appearing in the constitutive equations,
  so all the coefficients of the polynomials should be real.
\item
  The functions $L(z,\bm{w})$ and $M(\bm{w})$ are ``polynomials'',
  \thatis, the series are truncated to have a finite order of the derivatives
  so that the degrees of the polynomials $L(z,\bm{w})$ and $M(\bm{w})$
  with respect to $z$ and $\bm{w}$ are finite.
\item
  The constant term of each polynomial is
  the identity in the space of the dissipative currents,
  \thatis, $\Delta$ defined in Eq.~\eqref{eq:hydro.generic-dissipative-projector}:
  \begin{align}
    L(z,\bm{w}) &= \Delta + O(z,\bm{w}),
    \vv\label{eq:chf.general-polynomial-one-L} \\
    M(\bm{w}) &= \Delta + O(\bm{w}).
    \vv\label{eq:chf.general-polynomial-one-M}
  \end{align}
  This is because the lowest-order terms are
  $\Gamma = \kappa F(x)$.
\item
  The polynomials $L(z,\bm{w})$ and $M(\bm{w})$
  contain only even orders of $\bm{w}$:
  \begin{align}
    L(z,\bm{w}) &= L_2(z,\bm{w}\otimes\bm{w}),
      \vv\label{eq:chf.general-polynomial-even-L} \\
    M(\bm{w}) &= M_2(\bm{w}\otimes\bm{w}).
      \vv\label{eq:chf.general-polynomial-even-M}
  \end{align}
  This can be obtained by consideration
  on the parity inversion $\bm{x}\to\bm{x}'=-\bm{x}$.
  Because the constant terms of the polynomials are unity,
  the parity of the polynomials are $L(z,\bm{w})$ and $M(\bm{w})$ are $+1$.
  Note that, if the parity of the dissipative currents
  and the thermodynamic forces is different,
  the difference should be absorbed in the parity of the Onsager coefficients $\kappa$,
  so that the polynomials are always parity $+1$.
  Therefore, the polynomials are even as a function of $\bm{w}$.
  It should be noted that the notation $\bm{w}\otimes\bm{w}$ expresses
  that the even factors do not necessarily have the form of $\bm{w}\cdot\bm{w}$
  because the indices of $\bm{w}$ may be contracted with spatial indices of thermodynamic forces,
  e.g., $M(\nabla^\mu)\kappa \Gamma \sim \nabla^\alpha\nabla^\beta(2\eta\partial_{\langle\alpha}u_{\beta\rangle})$.
\end{itemize}
Then the Fourier representation of the differential form becomes
\begin{align}
  L(-i\omega,-i\bm{k}) \Gamma_{\omega,\bm{k}}
    &= M(-i\bm{k}) \kappa F_{\omega,\bm{k}} + \xi_{\omega,\bm{k}}.
  \vv\label{eq:fdr.differential-form.fourier}
\end{align}
Here $\D = \partial_0$ was replaced by $-i\omega$,
and $\nabla^\mu = (0, \partial^i) = (0, -\partial_i)$ was replaced by $-i\bm{k}$
since we now consider the coordinates in the rest frame.

Comparing the integral form~\eqref{eq:fdr.integral-form.fourier}
and the differential form~\eqref{eq:fdr.differential-form.fourier},
one finds the relations between the memory function and the polynomials
and between the noise terms of the two forms:
\begin{gather}
  G_{\omega,\bm{k}} = L(-i\omega,-i\bm{k})^{-1} M(-i\bm{k}),
  \vv\label{eq:fdr.memory-function-and-polynomials}\\
  \xi_{\omega,\bm{k}} = L(-i\omega,-i\bm{k}) \delta\Gamma_{\omega,\bm{k}}.
  \vv\label{eq:fdr.general-noise-relation}
\end{gather}

The FDR in the integral form~\ref{eq:fdr.uniform-fdr}
is transformed into the following form in the Fourier representation:
\begin{align}
  \langle \delta\Gamma_{\omega,\bm{k}}
          \delta\Gamma_{\omega',\bm{k}'}^\dag\rangle
    &= T(G_{\omega,\bm{k}}\kappa + \kappa G^\dag_{\omega,\bm{k}}) (2\pi)^4 \delta^{(4)}(k-k'),
    \vv\label{eq:chf.general-integral-noise-fdr}
\end{align}
where $\cdots^\dag$ denotes the Hermitian conjugate of a matrix or a vector.
The delta function $\delta^{(4)}(k-k')$ appears in the right-hand side
due to the translational symmetry of the memory function $G(x-x')$.
The FDR in the differential form is obtained as
\begin{align}
  \langle\xi_{\omega,\bm{k}}
        \xi^\dag_{\omega',\bm{k}'}\rangle
    &= L(-i\omega,-i\bm{k}) \langle \delta\Gamma_{\omega,\bm{k}}
          \delta\Gamma_{\omega',\bm{k}'}^\dag\rangle L(-i\omega',-i\bm{k}')^\dag \\
    &= I_{\omega, \bm{k}}(2\pi)^4 \delta^{(4)}(k-k'),
    \vv\label{eq:chf.general-differential-noise-fdr} \\
  I_{\omega,\bm{k}}
    &= T[M(-i\bm{k}) \kappa L(-i\omega,-i\bm{k})^\dag
      + L(-i\omega,-i\bm{k}) \kappa M(-i\bm{k})^\dag].
    \vv\label{eq:chf.general-differential-noise-fdr-power}
\end{align}
Here $I_{\omega,\bm{k}}$ represents the power spectrum of the noise field.
The noise is white if the power spectrum is constant,
and otherwise the noise is colored.
The expression of $I_{\omega,\bm{k}}$~\eqref{eq:chf.general-differential-noise-fdr-power}
in general does not have the same spectrum with the Navier--Stokes theory $I_{\omega,\bm{k}} = 2T\kappa$.
Also $I_{\omega,\bm{k}}$ is not white in general
because it depends on the frequency and the momentum $(\omega, \bm{k})$
through the polynomials $L(-i\omega,-i\bm{k})$ and $M(-i\bm{k})$.

However, there are additional physical conditions
which restrict the structure of the polynomials $L(z,\bm{w})$ and $M(\bm{w})$.
For example, the properties of the memory function
such as the causality~\eqref{eq:hydro.memory-function.causality},
the retardation~\eqref{eq:hydro.memory-function.retardation}
and the relaxation~\eqref{eq:hydro.memory-function.relaxation}
give constraints on the polynomials
through the relation~\eqref{eq:fdr.memory-function-and-polynomials}.
Also the autocorrelation $\langle\xi(x)\xi(x')^\T\rangle$
is a variance--covariance matrix when its dependence on $x$ and $x'$ is seen as matrix indices,
and therefore the autocorrelation should be positive semi-definite
due to the general property of a variance--covariance matrix.
This fact restricts the polynomials
through the expression of power spectrum~\eqref{eq:chf.general-differential-noise-fdr-power}
written in terms of the polynomials.

When the dissipative currents do not mix with one another,
all the matrices such as $\kappa$, $G_{\omega,\bm{k}}$, $L(-i\omega,-i\bm{k})$ and $M(-i\bm{k})$ are diagonal,
so that we can consider a single-component dissipative current for $\Gamma(x)$ without loss of generality.
In such a case, it can be shown that the restrictions on the polynomials
and the power spectrum have the following forms (see Appendix~\ref{sec:white}):
\begin{itemize}
\item
  The polynomial $L(\D,\nabla^\mu)$
  should have the following form:
  \begin{align}
    L(\D,\nabla^\mu) &= 1+\tau_R \D,
    \vv\label{eq:fdr.white.polynomial-L}
  \end{align}
  where the transport coefficient $\tau_R$ is
  identified to be the relaxation time of the dissipative current.
\item
  The polynomial $M(-i\bm{k})$
  is non-negative for all real $\bm{k}$:
  \begin{align}
    M(-i\bm{k})&\ge 0.
    \vv\label{eq:fdr.white.polynomial-M}
  \end{align}
\item
  The power spectrum of the noise in the differential form
  is independent of $\omega$,
  \thatis, the noise is white in the frequency space:
  \begin{align}
    I_{\omega,\bm{k}} &= 2T\kappa M(-i\bm{k}).
    \vv\label{eq:fdr.white.power-spectrum}
  \end{align}
\end{itemize}

From the power spectrum~\eqref{eq:fdr.white.power-spectrum},
the autocorrelation in the real space is obtained as
\begin{align}
  \langle\xi(x)\xi(x')\rangle
    &= 2T\kappa M(\nabla^\mu) \delta^{(4)}(x-x').
  \vv\label{eq:chf.general-result-differential-autocorrelation}
\end{align}
Similarly to the case of Eq.~\eqref{eq:fdr.uniform-IS.differential-fdr-diffusion},
it is subtle to interpret the derivatives on the delta function as the autocorrelation.
Thus the polynomial $M$ is further restricted to $M(\nabla^\mu) = 1$.
This means that the differential form of the constitutive equation
is restricted to the simplified Israel--Stewart theory
to be consistent with the noise terms obeying the FDR,
and the noise is white as already shown in Sec.~\ref{sec:fdr.simplified-IS}.

While a careful proof of these restrictions are given,
in Appendix~\ref{sec:white},
we here briefly outline the essential part of the proof
to use it in later discussions.
First we consider the positive semi-definiteness
which can be obtained from Eq.~\eqref{eq:chf.general-differential-noise-fdr-power} as follows:
\begin{align}
  I_{\omega,\bm{k}} = 2T\kappa \Re [M(-i\bm{k})L(-i\omega,-i\bm{k})] \ge 0.
  \label{eq:fdr.brief-proof.positive-semidefiniteness}
\end{align}
Here the polynomial $L(-i\omega,-i\bm{k})$ can be factorized with respect to $-i\omega$ as
\begin{align}
  L(-i\omega,-i\bm{k})
   &= (-1)^N A_{\bm{k}} \prod_{p=1}^N i(\omega - \omega_p(\bm{k})),
\end{align}
where $N = \deg_\omega L$
is the degree of $-i\omega$ in the polynomial $L(-i\omega,-i\bm{k})$.
The zeroes $\{\omega_p(\bm{k})\}_{p=1}^N$ corresponding to the poles of the memory function
must have negative imaginary part, \thatis, $\Im \omega_p(\bm{k}) < 0$,
to keep the memory function retarded. For this reason,
when $\omega$ moves from $-\infty$ to $\infty$,
the complex argument of the factor $\arg (\omega - \omega_p(\bm{k}))$
continuously decreases by the amount of $\pi$.
Then, the argument of polynomial $L(-i\omega, -i\bm{k})$ decreases by $N\pi$ in total.
Here, the change of the arguments is restricted within $\pi$
due to the inequality~\eqref{eq:fdr.brief-proof.positive-semidefiniteness}, which implies $N\le1$.
With additional consideration on symmetries and the causality,
the polymials are constrained
to the forms of Eqs.~\eqref{eq:fdr.white.polynomial-L}--\eqref{eq:fdr.white.polynomial-M}.

%%%%%%%%%%%%%%%%%%%%%%%%%%%%%%%%%%%%%%%%%%%%%%%%%%%%%%%%%%%%%%%%%%%%%%%%%%%%%%
\subsection{Cutoff scale and structure of differential form}
\vv\label{sec:cutoff-and-white-noise}

In Sec.~\ref{sec:fdr.simplified-IS}
we discussed that it is difficult to interpret the noise autocorrelation~\eqref{eq:fdr.uniform-IS.differential-fdr}
which contains the derivatives of delta functions
when the dissipative currents mix with one another through $[\tau_R \kappa]^\mathrm{A}$.
Also, in Sec.~\ref{sec:general-differential-form}, we argued that
even when the dissipative currents do not mix with one another,
the structure of the differential form is restricted
in Eqs.~\eqref{eq:fdr.white.polynomial-L}~and~\eqref{eq:fdr.white.polynomial-M}
due to the positive semi-definiteness of the autocorrelation,
the causality of the memory function, etc.
We also discussed that the polynomial $M(-i\bm{k})$ is further restricted to be unity
because of the derivatives of the delta function in the autocorrelation%
~\eqref{eq:chf.general-result-differential-autocorrelation}.
The differential forms which have derivatives of delta functions in its noise autocorrelation
or do not satisfy the restriction are not compatible with the naive FDR~\eqref{eq:fdr.uniform-fdr-theta}.

However, all of those discussions are based on the assumption that the FDR
is applicable to the noise with arbitrarily short scales.
For example, if an appropriate upper bound of the frequency $\omega$ or the wave numbers $\bm{k}$ is introduced,
the power spectrum of the noise is bounded
even if derivatives (which are translated to $-i\omega$ and $i\bm{k}$ in the spectrum)
are present in the front of the delta function.
For another example, when we considered the restriction to the polynomial $L(-i\omega,-i\bm{k})$
from the positive semi-definiteness of the autocorrelation,
we took the limits $\omega\to\pm\infty$ in Eq.~\eqref{eq:white.positive-semidefinite.arg-limit}.
We also considered the limit of $|\bm{k}|\to\infty$ in discussing the causality of the memory function
represented in the derivative expansion.
In the first place, the differential form is based on the derivative expansion
in which the derivatives are assumed to be small enough with respect to the expansion coefficients,
so it is subtle to discuss the short scale noise in the differential form.
In this sense, even the problem of the acausal modes in the Navier--Stokes theory
could be regarded as a kind of artifact of applying
the constitutive equation of the Navier--Stokes theory to arbitrarily small scales.

In fact, in non-linear fluctuating hydrodynamics,
one should introduce the cutoff length scale $\lambda_x$ and the cutoff time scale $\lambda_t$ of the hydrodynamic fluctuations.
Because the autocorrelation of the noise terms is given in terms of delta functions,
if infinitesimally small scales of the hydrodynamic fluctuations are considered,
the fluctuations of the hydrodynamic fields diverge to have unphysical values.
For example, the pressure can be negative due to infinitely large fluctuations.
In actual physical systems, there should be some coarse-graining scale
which gives a lower limit of the scale where hydrodynamics is applicable.
By introducing the hydrodynamic fluctuations,
such a lower limit of the scale could be lowered
so that the applicability of hydrodynamics is extended.
However, there should be still a non-zero lower limit of the scales
in which fluctuating hydrodynamics is applicable, such as $\lambda_x$ and $\lambda_t$.
Therefore in fluctuating hydrodynamics
only the hydrodynamic fluctuations of longer scales than $\lambda_x$ and $\lambda_t$
should be considered.

In such a framework,
the power spectrum of the noise~\eqref{eq:chf.general-differential-noise-fdr-power}
should be modified in the Fourier representation
so that the noise with smaller scales are suppressed.
While several ways of the regularization can be considered,
we here employ a sharp momentum cutoffs by the Heaviside functions:
\begin{align}
  I_{\omega,\bm{k}}
    &= T[M(-i\bm{k}) \kappa L(-i\omega,-i\bm{k})^\dag
      + L(-i\omega,-i\bm{k}) \kappa M(-i\bm{k})^\dag] \Theta(1 - \lambda_t \omega)\Theta(1 - \lambda_x\bm{|k|}).
\end{align}
With this regularized FDR, the power spectrum is bounded even when one takes the limit $|\omega|,|\bm{k}| \to \infty$.
In the position space, the delta functions are replaced by the Bessel functions coming from the Heaviside functions:
\begin{align}
  \langle\xi(x)\xi(x')\rangle
    &= T[M(\nabla^\mu) \kappa L(-\D,-\nabla^\mu)^\T + L(\D,\nabla^\mu) \kappa M(-\nabla^\mu)^\T]
    \frac{j_0(|t-t'|/\lambda_t)}{2\pi^3\lambda_t\lambda_x^3}
    \frac{j_1(|\bm{x}-\bm{x}'|/\lambda_x)}{|\bm{x}-\bm{x}'|/\lambda_x},
\end{align}
where $j_0(x)$ and $j_1(x)$ are the spherical Bessel functions of the order 0 and 1.
The derivatives on the Bessel functions can be safely interpreted as the autocorrelations.

With the cutoff scales, the restriction from the the positive semi-definiteness is also loosened.
The positive semi-definiteness could not be satisfied
when the degree of the polynomial $N = \deg_\omega L(-i\omega,-i\bm{k})$ is larger than one,
because its complex argument changes too much when one takes the high-frequency limit in
Eq.~\eqref{eq:white.positive-semidefinite.arg-limit}.
However, 
when one introduces a cutoff time scale $\lambda_t$,
the positive semi-definiteness of $\Re[M(-i\bm{k}) L(-i\omega,-i\bm{k})]$
is only required in the frequency region $|\omega| < 1/\lambda_t$
so that the differential form with $N\ge2$ can be valid with an appropriate cutoff scale.
Conversely, this means that the structure of the differential form
naturally introduces cutoff scales.
For example, let us consider the following differential form for a single-component dissipative current:
\begin{align}
  (1 + \tau_1 \D)(1 + \tau_2 \D) \Gamma(x) = \kappa F(x) + \xi(x),
\end{align}
where $\tau_1$ and $\tau_2$ are transport coefficients
which can be identified to be two different relaxation times
by calculating the memory function.
The power spectrum of the noise is obtained as
\begin{align}
  I_{\omega,\bm{k}}
    &= 2\kappa T \Re[(1 -i\omega\tau_1)(1 -i\omega\tau_2)] %\nonumber \\
    = 2\kappa T (1-\omega^2\tau_1\tau_2).
\end{align}
If one takes the limit $\omega\to\pm\infty$, one can see that the positive semi-definiteness is broken.
For the positive semi-definiteness of the spectrum, the frequency should have an upper bound:
$|\omega|\le (\tau_1\tau_2)^{-1/2}$.
Therefore the cutoff scales should be introduced as
\begin{align}
  I_{\omega,\bm{k}}
    &= 2\kappa T (1-\omega^2\tau_1\tau_2) \Theta(1 - \lambda_t \omega) \Theta(1 - \lambda_x\bm{|k|}),
\end{align}
where the cutoff time scale satisfies $\lambda_t \ge \sqrt{\tau_1\tau_2}$.
In this way, the structure of the differential form of the constitutive equation
naturally introduces the lower bound of the cutoff time or length scales.
The exceptions are the Navier--Stokes theory
and the simplified Israel--Stewart theory with non-mixing dissipative currents.
In these exceptional theories, the positive semi-definiteness of the noise power spectrum
is always satisfied without any cutoff scales.

%%%%%%%%%%%%%%%%%%%%%%%%%%%%%%%%%%%%%%%%%%%%%%%%%%%%%%%%%%%%%%%%%%%%%%%%%%%%%%%
\section{FDR in non-static and inhomogeneous backgrounds}
\vv\label{sec:inhomogeneous}

\subsection{Integral form FDR}
In Sec.~\ref{sec:fdr} we discussed the hydrodynamic fluctuations in equilibrium,
\thatis, the nature of the hydrodynamic fluctuations in static and homogeneous backgrounds.
While, in dynamical calculations of the high-energy nuclear collisions,
the matter with highly inhomogeneous profile and short lifetimes of the order $\sim \text{fm/$c$}$
should be described in relativistic hydrodynamics.
The matter created in high-energy nuclear collisions
basically expands in the direction of beam axis with relativistic velocities,
and also the matter is surrounded by the vacuum
so that near its boundary the temperature gradient becomes very large.
In such a situation, the FDR given in Sec.~\ref{sec:fdr}, Eq.~\eqref{eq:fdr.uniform-fdr-theta}, becomes subtle.
For example, the temperature $T$ can be in general different at two different points, $x$ and $x'$,
so it is non-trivial to determine
physically what temperature should be used to generate the hydrodynamic fluctuations.

The expression for the FDR in non-static and inhomogeneous backgrounds
can be found in the generalized Green--Kubo formula
for the Zubarev's non-equilibrium statistical operator~\cite{Zubarev:1974book,Hosoya:1983id},
in which the FDR for the integral form of the constitutive equation has the following form:
\begin{align}
  \langle\delta\Gamma(x) \delta\Gamma(x')^\T\rangle \Theta(x^0 - x'^0) &= G(x,x')\kappa(x') T(x'),
  \vv\label{eq:inhomogeneous.fdr-theta}
\end{align}
or equivalently,
\begin{align}
  \langle\delta\Gamma(x) \delta\Gamma(x')^\T\rangle
    &= G(x,x')\kappa(x') T(x') + T(x)\kappa(x) G(x',x)^\T.
  \vv\label{eq:inhomogeneous.fdr}
\end{align}

For the Navier--Stokes theory,
the FDR is unmodified from the equilibrium case
because the memory function is just a delta function:
\begin{align}
  \langle\delta\Gamma(x) \delta\Gamma(x')^\T\rangle
    &= 2T(x)\kappa(x)\delta^{(4)}(x-x').
  \vv\label{eq:inhomogeneous.fdr-NS}
\end{align}

For the simplified Israel--Stewart theory, the FDR is written
by the memory function in inhomogeneous backgrounds~\eqref{sec:pathline.simplified-IS.memory}--\eqref{sec:pathline.simplified-IS.memory-diffusion}:
\begin{align}
  \langle\Pi(x)\Pi(x')\rangle \Theta(x^0 - x'^0)
    &= \exp\left(-\int_{\tau(x')}^{\tau(x)} \frac{\d\tau_2}{\tau_\Pi(\tau_2,\bm{\sigma}(x))}\right)
    \frac{\zeta(x')T(x')}{\tau_\Pi(x')} \theta^{(4)}(\sigma(x), \sigma(x')),
    \vv\label{eq:inhomogeneous.fdr-IS-shear} \\
  \langle\pi^{\mu\nu}(x)\pi^{\alpha\beta}(x')\rangle \Theta(x^0 - x'^0)
    &= \exp\left(-\int_{\tau(x')}^{\tau(x)} \frac{\d\tau_2}{\tau_\pi(\tau_2,\bm{\sigma}(x))}\right)
      \frac{2\eta(x')T(x')}{\tau_\pi(x')} \nonumber \\ & \qquad \times
      \Delta(\tau(x);\tau(x'),\bm{\sigma}(x))^{\mu\nu\alpha\beta} \theta^{(4)}(\sigma(x), \sigma(x')),
    \vv\label{eq:inhomogeneous.fdr-IS-bulk} \\
  \langle\nu_i^\mu(x)\nu_j^\alpha(x')\rangle \Theta(x^0 - x'^0)
    &= \sum_{k,l=1}^n
      \left[\T\exp\left(-\int_{\tau(x')}^{\tau(x)} \d\tau_2 \tau^{-1}_{ik}(\tau_2,\bm{\sigma}(x))\right)\right]_{ik}
      \tau_{kl}^{-1}(x')\kappa_{lj}(x') T(x') \nonumber \\ & \qquad \times
      [-\Delta(\tau(x);\tau(x'),\bm{\sigma}(x))^{\mu\alpha}] \theta^{(4)}(\sigma(x), \sigma(x')).
    \vv\label{eq:inhomogeneous.fdr-IS-diffuse}
\end{align}

For the general linear-response differential form,
due to the restriction~\eqref{eq:fdr.white.polynomial-L},
the constitutive equations have
the form~\eqref{eq:pathline.memfun.ce-X.shear} and \eqref{eq:pathline.memfun.ce-X.diffuse}
with $X^{\mu\nu} = M(\nabla^\mu)\ud{\mu\nu}{\alpha\beta} 2\eta \partial^{\langle\alpha}u^{\beta\rangle}$
and $X^\mu = M(\nabla^\mu)\ud\mu\alpha T\kappa_{ij}\nabla^\alpha \frac{\mu_j}{T}$.
Here, to apply the restriction,
we assumed that the diffusion currents do not mix with one another.
The components of the shear stress tensor essentially
do not mix with one another due to the rotational symmetry.
Therefore the corresponding integral forms can be written in terms of the pathline projectors as in
Eqs.~\eqref{eq:pathline.memfun.ce-X-solution.shear}--\eqref{eq:pathline.memfun.ce-X-mem.diffuse}.
The FDR can be written down similarly
by the memory function extracted from these integral forms.
The FDR for the bulk pressure can also be written down similarly
but more simply without a pathline projector.

\subsection{Differential form FDR}
In dynamical calculations we apply
the differential form of the constitutive equation
to inhomogeneous and non-static fluid fields.
Therefore we need to consider the FDR
in the differential form in inhomogeneous backgrounds.
Let us start from the general linear constitutive equation~\eqref{eq:fdr.higher-order.differential-form}
in inhomogeneous backgrounds:
\begin{align}
  L(\D,\nabla^\mu; x) \Gamma(x) &= M(\nabla^\mu; x) \kappa F(x) + \xi(x).
  \vv\label{eq:inhomogeneous.general-differential-form}
\end{align}
Here it should be noted that the coefficients of polynomials,
$L$ and $M$, now have spatial dependence on $x$
unlike in the static and homogeneous backgrounds.
Also, comparing the integral form~\eqref{eq:fdr.integral-form-with-noise}
and Eq.~\eqref{eq:inhomogeneous.general-differential-form},
one finds relations of memory functions, polynomials and the noise terms,
which correspond to the equilibrium version~%
\eqref{eq:fdr.memory-function-and-polynomials} and \eqref{eq:fdr.general-noise-relation}:
\begin{gather}
  L(\D,\nabla^\mu; x) G(x, x') = M(\nabla^\mu; x) \delta^{(4)}(x-x'), \\
  \xi(x) = L(\D,\nabla^\mu; x) \delta\Gamma(x).
\end{gather}

The autocorrelation of the noise in the differential form is calculated as
\begin{align}
  \langle\xi(x)\xi(x')^\T\rangle
  &= L(\D,\nabla^\mu;x)
    \langle \delta\Gamma(x)\delta\Gamma(x')^\T\rangle
    L(\overleftarrow{\D}', \overleftarrow{\nabla}'^\mu; x')^\T \nonumber \\
  &= L(\D,\nabla^\mu;x)
    [ G(x,x')\kappa(x') T(x') + T(x)\kappa(x) G(x',x)^\T ]
    L(\overleftarrow{\D}', \overleftarrow{\nabla}'^\mu; x')^\T \nonumber \\
  &= M(\nabla^\mu; x)\delta^{(4)}(x-x')\kappa(x')T(x')L(\overleftarrow{\D}', \overleftarrow{\nabla}'^\mu; x')^\T
    \nonumber \\ & \quad
   + L(\D,\nabla^\mu;x) T(x)\kappa(x) \delta^{(4)}(x-x')M(\overleftarrow{\nabla}'^\mu; x')^\T,
  \vv\label{eq:inhomogeneous.differential-fdr-general}
\end{align}
where $\overleftarrow{\D}'$ and $\overleftarrow{\nabla}'$ are the derivatives
with respect to $x'$
that operate on the left-hand side of the polynomials $L$ and $M$.
Note that the derivatives do not operate on the coefficients in the polynomials.

For the Navier--Stokes theory in which $L = M = \Delta$,
the FDR is reduced to the normal one
as is already shown in Eq.~\eqref{eq:inhomogeneous.fdr-NS}.

For the simplified Israel--Stewart theory
in which $L = \Delta + \tau_R \mathcal{D} = \Delta (\identity + \tau_R \D)$ and $M = \Delta$,
the FDR is calculated as follows:
\begin{align}
  \langle\xi(x)\xi(x')^\T\rangle
  &= \Delta(x)\delta^{(4)}(x-x')\kappa(x')T(x')[\identity + \overleftarrow{\D}' \tau_R^\T(x')] \Delta(x')^\T
    \nonumber \\ &\qquad
   + \Delta(x)[\identity + \tau_R(x)\D] T(x)\kappa(x) \delta^{(4)}(x-x')\Delta(x')^\T \nonumber \\
  &= 2\{T(x)\kappa(x) + \Delta(x) [\tau_R(x) \D T(x)\kappa(x)]^{\mathrm{S}}\Delta(x)^\T \} \delta^{(4)}(x-x')
    \nonumber \\ & \quad
    + \Delta(x) [ K(x')^\T \D'\delta^{(4)}(x-x')
    + K(x) \D \delta^{(4)}(x-x') ] \Delta(x')^\T,
\end{align}
where $K(x) \eqdef \tau_R(x)T(x)\kappa(x)$ is defined,
$A^{\mathrm{S}} \eqdef (A + A^\T)/2$ is the symmetric part of a matrix $A$,
and the relation $\Delta(x) \kappa(x) = \kappa(x) \Delta(x)^\T = \kappa(x)$ is used.
The symmetric part of the last line can be transformed as follows:
\begin{align}
  & K(x')^{\mathrm{S}}\D'\delta^{(4)}(x-x') + K(x)^{\mathrm{S}}\D\delta^{(4)}(x-x') \nonumber \\
  & \quad = \partial'_\mu[u(x')^\mu K(x')^{\mathrm{S}}\delta^{(4)}(x-x')]
    - [\partial'_\mu u(x')^\mu K(x')^{\mathrm{S}}]\delta^{(4)}(x-x')
    + K(x)^{\mathrm{S}}\D\delta^{(4)}(x-x') \nonumber \\
  & \quad = -K(x)^{\mathrm{S}}\D\delta^{(4)}(x-x') - [\D K(x)^{\mathrm{S}} + \theta(x)K(x)^{\mathrm{S}}]\delta^{(4)}(x-x') + K(x)^{\mathrm{S}}\D\delta^{(4)}(x-x') \nonumber \\
  & \quad = - [\D K(x)^{\mathrm{S}} + \theta(x)K(x)^{\mathrm{S}}]\delta^{(4)}(x-x').
\end{align}
where the relation $\partial_{y}[f(y)\delta(x-y)] = \partial_{y}[f(x)\delta(x-y)] = f(x) \partial_{y}\delta(x-y) = -f(x)\partial_x \delta(x-y)$
is used to obtain the third line.
Here it should be noted that the derivatives on the delta functions should be treated carefully.
For example, the relation $\partial_x\delta(x) = -\delta(x)\partial_x$ does not apply to $\delta(x-y)$ when $y$ is not fixed,
\thatis, $\partial_x\delta(x-y) \neq -\delta(x-y)\partial_x$.
This is because
\begin{align}
  \int_D \d{x} f(x) \partial_x \delta(x-y) = \left.[f(x)\delta(x-y)]\right|_{\partial D} - \int_D \d{x} \delta(x-y) \partial_x f(x),
\end{align}
where the surface term does not vanish when $y$ is on the boundary.
Finally one obtains the following differential FDR:
\begin{align}
  \langle\xi(x)\xi(x')^\T\rangle
  &= 2T\kappa \delta^{(4)}(x-x')
    \nonumber \\ & \quad
    + \Delta(x) [ K(x) \D \delta^{(4)}(x-x') - K(x') \D'\delta^{(4)}(x-x') ]^{\mathrm{A}} \Delta(x')^\T
    \nonumber \\ & \quad
    + \Delta [\tau_R\D T\kappa - (\D \tau_R)T\kappa - \theta \tau_R T \kappa]^{\mathrm{S}}\Delta^\T \delta^{(4)}(x-x'),
    \vv\label{eq:inhomogeneous.IS-differential-fdr-general}
\end{align}
where the spatial dependence on $x$ or $x'$ of the coefficients of the first and third terms
is omitted because it does not matter due to the delta function.
Here the first and second term correspond
to the terms in the FDR in equilibrium~\eqref{eq:fdr.uniform-IS.differential-fdr}.
The second term is slightly modified but is easy to check that in equilibrium
it exactly reduces to the second term of Eq.~\eqref{eq:fdr.uniform-IS.differential-fdr}.
As in the equilibrium case, this antisymmetric part vanishes
when one considers the dissipative currents not mixing with one another.
The third term is a newly obtained modification to the FDR
originating from the time evolution of the background.
The FDR for each dissipative current reads
\begin{align}
  \langle\xi_\Pi(x)\xi_\Pi(x')\rangle
    &= \left(2 + \tau_\Pi\D\ln\frac{T\zeta}{\tau_\Pi} - \tau_\Pi\theta \right) T\zeta \delta^{(4)}(x-x'),
    \vv\label{eq:inhomogeneous.differential-fdr-bulk} \\
  \langle\xi_\pi^{\mu\nu}(x)\xi_\pi^{\alpha\beta}(x')\rangle
    &= 2\left[
      \left(2 + \tau_\pi\D\ln\frac{T\eta}{\tau_\pi} - \tau_\pi\theta \right)\Delta^{\mu\nu\alpha\beta}
      +\tau_\pi\mathcal{D}\Delta^{\mu\nu\alpha\beta} \right]
      T\eta \delta^{(4)}(x-x'), \\
  \langle\xi_i^\mu(x)\xi_j^\alpha(x')\rangle
    &= -2T\kappa_{ij} \Delta^{\mu\alpha} \delta^{(4)}(x-x')
     \nonumber \\ & \quad
      - \Delta^{\mu\alpha} [K_{ij}^{\mathrm{A}}(x)\mathcal{D} - K_{ij}^{\mathrm{A}}(x') \mathcal{D}'] \delta^{(4)}(x-x')
    \nonumber \\ & \quad
      +\sum_{k=1}^n \left\{ -\Delta^{\mu\alpha} \left[
        \tau_{ik} \D T \kappa_{kj} - (\D\tau_{ik})T\kappa_{kj} - \tau_{ik}\theta T \kappa_{kj} \right]^{\mathrm{S}}
        -K_{ij}^\mathrm{S} \mathcal{D} \Delta^{\mu\alpha} \right\}
      \delta^{(4)}(x-x'),
    \vv\label{eq:inhomogeneous.differential-fdr-diffuse}
\end{align}
where $K_{ij}^{\mathrm{S}/\mathrm{A}}(x) = \sum_{k=1}^n T(x)(\tau_{ik}(x) \kappa_{kj}(x) \pm \tau_{jk}(x) \kappa_{ki}(x)) / 2$,
and $[\circ_{ij}]^\mathrm{S} = (\circ_{ij} + \circ_{ji})/2$.
The derivative $\mathcal{D}$ on the projectors is defined as
\begin{align}
  \mathcal{D}\Delta^{\mu\nu\alpha\beta}
    &= \Delta\ud{\mu\nu}{\kappa\lambda}\Delta\ud{\alpha\beta}{\gamma\delta} \D \Delta^{\kappa\lambda\gamma\delta}, \\
  \mathcal{D}\Delta^{\mu\alpha}
    &= \Delta\ud{\mu}{\kappa}\Delta\ud{\alpha}{\gamma} \D \Delta^{\kappa\gamma}.
\end{align}

As already seen in the bulk pressure case \eqref{eq:inhomogeneous.differential-fdr-bulk},
the typical structure of the FDR for a single-component dissipative current is
\begin{align}
  \langle\xi(x)\xi(x')\rangle
    &= \left(2 + \tau_R\D\ln\frac{T\kappa}{\tau_R} - \tau_R\theta \right) T\kappa \delta^{(4)}(x-x').
\end{align}
Because the expansion rate $\theta$ can be expressed
in terms of the fluid element volume $\Delta V$ as
$\theta = \D \ln \Delta V$,
the first factor can be regarded as $[2 + \tau_R\D\ln(T\kappa/\tau_R\Delta V)]$ intuitively.
This factor becomes negative
when $T\kappa/\tau_R\Delta V$ rapidly decreases in a shorter time scale than $\tau_R/2$,
which occurs, e.g., when the expansion rate becomes extremely large or the temperature suddenly decreases.
In such a situation, fluctuating hydrodynamics with the differential form of the constitutive equation
breaks down because the noise with a negative norm is unphysical.
Even if fluctuating hydrodynamics does not break down,
it is still important in the dynamical description of the high-energy nuclear collisions
because the matter created in the collision reaction
expands with the relativistic velocity in the longitudinal direction
so that $\theta$ is expected to be large,
and also because the lifetime of the matter is several $\text{fm}/c$
so that the temperature decreases in a very small time scale.

%%%%%%%%%%%%%%%%%%%%%%%%%%%%%%%%%%%%%%%%%%%%%%%%%%%%%%%%%%%%%%%%%%%%%%%%%%%%%%%
\section{Summary}
\label{sec:summary}

To describe the rapid spacetime evolution of the highly inhomogeneous matter
created in high-energy nuclear collisions,
non-linear equations of causal dissipative hydrodynamics
with a non-vanishing relaxation time should be solved in dynamical models.
In causal dissipative hydrodynamics,
the differential form of the constitutive equation,
in which the dissipative currents are treated as additional dynamical fields,
is used for the analytic investigations and numerical simulations
because of its simplicity and smaller computational complexity compared to the integral form.
To introduce the hydrodynamic fluctuations in such a framework,
we need to consider the non-trivial properties of hydrodynamic fluctuations
in non-static and inhomogeneous matter.
The autocorrelations of the hydrodynamic fluctuations are determined by the FDR,
which is normally based on the linear-response theory
of small perturbations to global equilibrium.
However, the background fields are no longer static and homogeneous
in hydrodynamic description of the created matter.
Even when the background fields are non-static and inhomogeneous,
one could just apply the FDR of global equilibrium
with the local thermal state given by the local temperature and chemical potentials
if the autocorrelations were local.
However, because of the relaxation time which is needed to maintain the causality,
the hydrodynamic fluctuations have non-local correlations
so that one cannot assume a single thermal state for the FDR\@.
Here we need to explicitly consider the FDR
in the non-static and inhomogeneous background
by considering the linear-response constitutive equations.

In Sec.~\ref{sec:hydro},
the differential and integral forms of the constitutive equation were explained.
For practical purpose,
we usually use the differential form of the constitutive equations~\eqref{eq:hydro.dissipative-deriv},
in which the dissipative current, $\Gamma(x)$, is represented
by the gradient expansion truncated to a finite order.
For examples, the differential forms
in the Navier--Stokes theory~\eqref{eq:hydro.navier-stokes} (which is first-order and acausal) and
the simplified Israel--Stewart theory~\eqref{eq:hydro.simplified-IS} (which is second-order and causal) are introduced.
While, in the linear-response theory,
the constitutive equation is naturally given by the integral form~\eqref{eq:hydro.integral-form},
in which the dissipative currents are given by past thermodynamic forces, $\kappa F(x')$,
convoluted with the integration kernel, $G(x,x')$, called the memory function.
The FDR from the linear-response theory is naturally given in this integral form.

The differential form of the constitutive equation
usually contains the time derivatives of the dissipative current
and therefore is an implicit form with respect to the dissipative current.
By solving the differential form with respect to the dissipative current,
the corresponding integral form shall be obtained for the later discussion of the FDR\@.
In global equilibrium where the translational symmetry can be utilized,
such a solution can be easily obtained in the Fourier representation.
However, it is non-trivial to obtain the solution
in non-static and inhomogeneous backgrounds.
In particular, the dissipative currents such as the shear stress and the diffusion currents
have special constraints such as the transversality to the flow velocity.
To write down the explicit form of the integral form
corresponding to a certain class of the differential forms
in non-static and inhomogeneous backgrounds,
we newly introduced in Sec.~\ref{sec:pathline} the pathline projectors,
$\Delta(\tau;\tau')\ud\mu\alpha$~\eqref{eq:pathline.def3}
and $\Delta(\tau;\tau')\ud{\mu\nu}{\alpha\beta}$~\eqref{eq:pathline.def5},
which perform projection to the tensor components of the dissipative currents
at every moment along a pathline.
Using the properties of the pathline projectors,
Eqs.~\eqref{eq:ttproj3.prop-repr}--\eqref{eq:ttproj5.prop-preserve-norm},
the integral form corresponding to the differential form~\eqref{eq:pathline.memfun.ce-X.shear}--\eqref{eq:pathline.memfun.ce-X.diffuse}
was obtained as in Eqs.~\eqref{eq:pathline.memfun.ce-X-solution.shear}--\eqref{eq:pathline.memfun.ce-X-mem.diffuse}.
In particular, for the simplified Israel--Stewart theory,
the explicit form of the memory function~\eqref{sec:pathline.simplified-IS.memory}--\eqref{sec:pathline.simplified-IS.memory-diffusion}
was obtained.

In Sec.~\eqref{sec:fdr},
we next discussed the FDR in higher-order linear-response constitutive equations in equilibrium.
First we introduced the noise term, $\delta\Gamma(x)$, in the integral form~\eqref{eq:fdr.integral-form-with-noise}
and gave its autocorrelation, $\langle\delta\Gamma(x)\delta\Gamma(x')\rangle$, by the FDR~\eqref{eq:fdr.uniform-fdr}.
In the Navier--Stokes theory the noise is white as usual,
\thatis, the noise autocorrelation is a delta function and has no characteristic frequency (or ``color'').
While, in causal theories with a non-vanishing relaxation time, $\tau_R$,
the noise in the integral form becomes colored,
\thatis, the autocorrelation has a characteristic ``color'' of $\sim 1/\tau_R$.
In the simplified Israel--Stewart theory,
the corresponding noise term, $\xi(x)$, appears in the differential form,
and its autocorrelation is given by
Eqs.~\eqref{eq:fdr.uniform-IS.differential-fdr}--\eqref{eq:fdr.uniform-IS.differential-fdr-diffusion}.
However, when the components of the dissipative current mix with one another,
the autocorrelation contains the derivative on the delta function which is naively unphysical.
For the general linear-response differential form~\eqref{eq:fdr.higher-order.differential-form},
the noise autocorrelation is given by
Eqs.~\eqref{eq:chf.general-differential-noise-fdr}--\eqref{eq:chf.general-differential-noise-fdr-power}.
When the dissipative currents do not mix with one another,
it was shown that the differential form is restricted to a very simple form with
Eqs.~\eqref{eq:fdr.white.polynomial-L}--\eqref{eq:fdr.white.polynomial-M}
using the positive semi-definiteness of
the noise autocorrelations~\eqref{eq:white.positive-semidefinite} and
the general properties of the memory function
such as the relaxation~\eqref{eq:hydro.memory-function.relaxation}
and the causality~\eqref{eq:hydro.memory-function.causality}.
Moreover, if one does not allow the derivative on the delta function,
the only allowed causal differential form is the simplified Israel--Stewart theory.
All of these restrictions come from the behavior of the hydrodynamics fluctuations
in infinitesimally small length and time scales.
However, in actual non-linear fluctuating hydrodynamics,
the cutoff scale of the hydrodynamic fluctuations should be introduced.
We discussed in Sec.~\ref{sec:cutoff-and-white-noise} that those restrictions can be
reinterpreted as the lower bound of the cutoff scales.

Finally, we discussed the FDR in inhomogeneous background given by Eq.~\eqref{eq:inhomogeneous.fdr}.
While the noise autocorrelation in the Navier--Stokes theory is not modified~\eqref{eq:inhomogeneous.fdr-NS},
the noise autocorrelation in the integral form of the simplified Israel--Stewart theory
is given by Eqs.~\eqref{eq:inhomogeneous.fdr-IS-shear}--\eqref{eq:inhomogeneous.fdr-IS-diffuse}
using the pathline projectors.
For the general linear-response differential form in inhomogeneous background~\eqref{eq:inhomogeneous.general-differential-form},
the noise autocorrelation is given by Eq.~\eqref{eq:inhomogeneous.differential-fdr-general}.
For the simplified Israel--Stewart theory,
we obtained the explicit form
of the FDR~\eqref{eq:inhomogeneous.IS-differential-fdr-general}--\eqref{eq:inhomogeneous.differential-fdr-diffuse}
to find that new modification terms appear and are proportional
to the relaxation time and the time derivatives of the background thermodynamic quantities.

The hydrodynamic fluctuations should be properly implemented
in dynamical models for the high-energy nuclear collisions
to satisfy the correct FDR
corresponding to the actually used differential form.
Also, the cutoff scales of the hydrodynamic fluctuations should be chosen
so that the positive-semidefiniteness of the noise autocorrelation is satisfied
when one uses the differential form other than the Israel--Stewart theory.
The effects of the modification to the FDR in non-static and inhomogeneous backgrounds
to the experimental observables have not yet been investigated.
They should also be implemented in dynamical models
and quantified by comparing the results to those without the modification
by performing event-by-event simulations.
Another future challenge would be to investigate the effects on the noise statistics coming from non-linear responses.
The second-order constitutive equations actually used
in the high-energy nuclear collisions contain the non-linear terms
with respect to the thermodynamic forces.
If one assumes that the non-linear terms are not too large compared to the linear terms
near the equilibrium, one could approximate the noise autocorrelation
by the FDR based on only the linear terms in the differential form.
However, the non-linear terms can have the same order
with the other second-order terms in actual dynamics.
They may cause non-Gaussian statistics of the fluctuations
and also may introduce additional modifications to the FDR\@.

\section*{Acknowledgments}
The author thanks Tetsufumi Hirano and Yuji Hirono for useful discussions
in the early stage of this study. This work was supported
by JSPS KAKENHI Grant Number 12J08554 in the early stage.

%%%%%%%%%%%%%%%%%%%%%%%%%%%%%%%%%%%%%%%%%%%%%%%%%%%%%%%%%%%%%%%%%%%%%%%%%%%%%%%
\appendix
\section{Properties of projectors}
\label{app:pathline.projector-property}
The properties of the projectors,
which can be directly shown by the definition of the projectors,
are listed here.
\begin{align}
  \Delta\ud\mu\kappa \Delta\ud\kappa\alpha &= \Delta\ud\mu\alpha,
    \vv\label{eq:pathline.proj-prop-proj3} \\
  \Delta\ud{\mu\nu}{\kappa\lambda} \Delta\ud{\kappa\lambda}{\alpha\beta} &= \Delta\ud{\mu\nu}{\alpha\beta},
    \vv\label{eq:pathline.proj-prop-proj5} \\
  \Delta^{\mu\nu} &= \Delta^{\nu\mu},
    \vv\label{eq:pathline.proj-prop-sym3} \\
  \Delta^{\mu\nu\alpha\beta} &= \Delta^{\alpha\beta\mu\nu}
    = \Delta^{\nu\mu\alpha\beta} = \Delta^{\mu\nu\beta\alpha},
    \vv\label{eq:pathline.proj-prop-sym5} \\
  \Delta\ud\mu\alpha u^\alpha &= 0,
    \vv\label{eq:pathline.proj-prop-trans3} \\
  \Delta\ud{\mu\nu}{\alpha\beta} u^\alpha &= 0,
    \vv\label{eq:pathline.proj-prop-trans5} \\
  \Delta\ud{\mu\nu}{\alpha\beta} \Delta^{\alpha\beta} &= \Delta\ud{\mu\nu}{\alpha\beta} g^{\alpha\beta} = \Delta\ud{\mu\nu\alpha}{\alpha} = 0,
    \vv\label{eq:pathline.proj-prop-traceless} \\
  \Delta\ud{\mu\nu}{\kappa\beta} \Delta\ud{\kappa}{\alpha} &= \Delta\ud{\mu\nu}{\alpha\beta}.
    \vv\label{eq:pathline.proj-prop-spatial}
\end{align}

%%%%%%%%%%%%%%%%%%%%%%%%%%%%%%%%%%%%%%%%%%%%%%%%%%%%%%%%%%%%%%%%%%%%%%%%%%%%%%%
\section{Proof of the convergence and the properties of the pathline projectors: Eqs.~\eqref{eq:ttproj3.prop-repr}--\eqref{eq:ttproj5.prop-dyn2}}
\label{sec:pathline.proof}

In later sections we will utilize the pathline projectors which have the useful properties,
Eqs.~\eqref{eq:ttproj3.prop-repr}--\eqref{eq:ttproj5.prop-dyn2}.
However these properties are non-trivial from the definition of the pathline projectors,
Eqs.~\eqref{eq:pathline.def3}~and~\eqref{eq:pathline.def5}.
In this section we aim to give proofs to these properties from the definition of the pathline projectors.
We give a proof to the convergence of the limits in Eqs.~\eqref{eq:pathline.def3}~and~\eqref{eq:pathline.def5},
and then give proofs to the properties.
The outline of the proof is as follows:
First we introduce a general projector $P(\tau)$ as a matrix-valued function of $\tau$,
and parametrize the projector by bases of $\ker P(\tau)$, such as $\{\bm{u}_i(\tau)\}_i$ and $\{\bm{v}_i(\tau)\}_i$,
which are assumed to be continuously differentiable for two times with respect to $\tau$, \thatis, $C^2$ functions of $\tau$.
We also define a function sequence $P_N(\tauF;\tauI)$ indexed by $N$.
Next we consider the derivative of $P_N(\tauF;\tauI)$
to identify the residual contribution $(1/N) R_N(\tauF;\tauI)$ to the derivative,
and show the boundedness of $R_N(\tauF;\tauI)$.
Finally we show the compact convergence of $P_N(\tauF;\tauI)$ and $\Df P_N(\tauF;\tauI)$
to obtain the properties of the pathline projectors.

It should be noted in advance that
the mean-value theorem and the Taylor's theorem will be repeatedly used in the proof:
for any $C^2$ function $f(x)$, there exist $\xi, \xi' \in I(x_0,x_1)$ such that
\begin{align}
  f(x_1) &= f(x_0) + (x_1-x_0) f'(\xi), \\
  f(x_1) &= f(x_0) + (x_1-x_0) f'(x_0) + \frac{(x_1-x_0)^2}2 f''(\xi'),
\end{align}
where the closed interval $I(x,y)$ is defined as $I(x,y) \eqdef \{z \;|\; \min\{x,y\} \le z \le \max\{x,y\} \}$.
Also, in this section dots are used to denote the time derivatives,
\thatis, $\dot f(\tau) = \D f(\tau)$, $\ddot f(\tau) = \D^2 f(\tau)$, etc.

\subsection{Definition of $P_N(\tauF;\tauI)$}
Let us consider an $n$-dimensional linear space, $V$,
and a time-dependent projector in $V$, $P(\tau)$,
which satisfies $P(\tau)^2 = P(\tau)$.
Using the right and left eigenvectors belonging to the eigenspace $\ker P(\tau)$,
$\{\bm{u}_i(\tau)\}_i$ and $\{\bm{v}_i(\tau)^\T\}_i$ normalized as $\bm{v}_i^\T\bm{u}_j = \delta_{ij}$,
the projector can be expressed as
\begin{align}
  P(\tau) &= \mathbbm{1} - \sum_{i=1}^{m} \bm{u}_i(\tau) \bm{v}_i(\tau)^\T,
\end{align}
where $\mathbbm{1}$ denotes the identity matrix in $V$, and $m \eqdef \dim \ker P(\tau)$.
Here it is assumed that $\bm{u}_i(\tau)$ and $\bm{v}_i(\tau)^\T$ are $C^2$ functions,
and the derivatives are bounded in the considered domain $D$:
\begin{align}
  M_p &\eqdef \sup_{i,a,\tau} \{|\D^p u_{ia}(\tau)|, |\D^p v_{ia}(\tau)|\} < \infty, \quad (p = 0, 1, 2),
\end{align}
where $u_{ia}(\tau) \eqdef \bm{e}_a^\T\bm{u}_i(\tau)$
and $v_{ia}(\tau) \eqdef \bm{v}_i(\tau)^\T \bm{e}_a$ ($a=1,\dots,n$)
are the $a$-th components of $\bm{u}_i(\tau)$ and $\bm{v}_i(\tau)^\T$, respectively,
with $\{\bm{e}_a\}_a$ being a (time-independent) orthonormal basis of $V$.
It should be noted that, for the projectors $\Delta\ud\mu\alpha$ and $\Delta\ud{\mu\nu}{\alpha\beta}$,
one can explicitly construct $\bm{u}_i$ and $\bm{v}_i$ in terms of $u^\mu$ (see Appendix~\ref{sec:pathline.appendix}).
Next a function sequence $\{P_N(\tauF; \tauI)\}_N$ is defined:
\begin{align}
  P_N(\tauF;\tauI) &\eqdef \begin{cases}
    \mathbbm{1} & (N < 0), \\
    P(\tauI) & (N = 0), \\
    \prod_{k=0}^{N} P(\tau_k) & (N > 0),
  \end{cases}
\end{align}
where $\tau_k \eqdef \tauF - \frac{\Delta\tau}{N} k$ and $\Delta\tau \eqdef \tauF - \tauI$.
The following properties can be easily shown:
\begin{align}
  P_N(\tauF;\tauI) &= P(\tauF)P_N(\tauF;\tauI) = P_N(\tauF;\tauI)P(\tauI),
    \vv\label{eq:proof.seqprop.1} \\
  P_N(\tauI;\tauI) &= P(\tauI),
    \vv\label{eq:proof.seqprop.2} \\
  P_N(\tauF;\tauI) &= P_N(\tauI;\tauF)^\T.
    \vv\label{eq:proof.seqprop.3}
\end{align}

\subsection{Derivative of $P_N(\tauF;\tauI)$}
The derivative of the single projector, $\dot P(\tau)$, can be calculated as
\begin{align}
  \dot P(\tau) &= -\sum_{i=1}^m [\dot{\bm{u}}_i(\tau)\bm{v}_i(\tau)^\T + \bm{u}_i(\tau)\dot{\bm{v}}_i(\tau)^\T],
  \vv\label{eq:proof.single-derivative.1} \\
  \dot P(\tau) &= \D[P(\tau)^2] = P(\tau) \dot P(\tau) + \dot P(\tau) P(\tau) \nonumber \\
    &= -\sum_{i=1}^m [P(\tau) \dot{\bm{u}}_i(\tau)\bm{v}_i(\tau)^\T + \bm{u}_i(\tau)\dot{\bm{v}}_i(\tau)^\T P(\tau)],
\end{align}
where the relation $P(\tau)\bm{u}_i(\tau) = \bm{v}_i(\tau)^\T P(\tau) = 0$ is used to obtain the last line.
For $N\ge2$, the derivative of a sequence term, $P_N(\tauF;\tauI)$, with respect to $\tauF$ is
\begin{align}
  \Df P_N(\tauF;\tauI)
  &= -\sum_{k=0}^{N-1} (1-\tfrac{k}{N}) P_{k-1}(\tauF;\tau_{k-1})
    \dot P(\tau_k)
    P_{N-k-1}(\tau_{k+1};\tauI) \nonumber \\
  &= -\sum_{k=0}^{N-1} (1-\tfrac{k}{N}) P_k(\tauF;\tau_k)
    \sum_{i=1}^m \dot{\bm{u}}_i(\tau_k)\bm{v}_i(\tau_k)^\T
    P_{N-k-1}(\tau_{k+1};\tauI) \nonumber \\ & \quad
  -\sum_{k=0}^{N-1} (1-\tfrac{k}{N}) P_{k-1}(\tauF;\tau_{k-1})
    \sum_{i=1}^m  \bm{u}_i(\tau_k)\dot{\bm{v}}_i(\tau_k)^\T
    P_{N-k}(\tau_k;\tauI)
    \vv\label{eq:proof.seq.deriv.2} \\
  &= -\sum_{k=1}^{N-1} P_{k-1}(\tauF;\tau_{k-1})
    \sum_{i=1}^m \left[
      (1-\tfrac{k-1}{N}) \dot{\bm{u}}_i(\tau_{k-1})\bm{v}_i(\tau_{k-1})^\T
      + (1-\tfrac{k}{N}) \bm{u}_i(\tau_k)\dot{\bm{v}}_i(\tau_k)^\T
    \right] P_{N-k}(\tau_k;\tauI) \nonumber \\ & \quad
  - \frac1N P_{N-1}(\tauF;\tau_{N-1})
    \sum_{i=1}^m \dot{\bm{u}}_i(\tau_{N-1})\bm{v}_i(\tau_{N-1})^\T
    P(\tauI)
  - \sum_{i=1}^m \bm{u}_i(\tauF)\dot{\bm{v}}_i(\tauF)^\T
    P_N(\tauF;\tauI),
    \vv\label{eq:proof.seq.deriv.3}
\end{align}
where $\Df \eqdef \partial/\partial\tauF$.
In the first line, the relation $\Df P(\tau_k) = (1- \tfrac{k}{N})\dot P(\tau_k)$ is used.
The index $k \to k' = k+1$
of the first summation of Eq.~\eqref{eq:proof.seq.deriv.2}
was shifted to obtain the last line.
From the Taylor's theorem,
there exist $\xi^{(1)}_{kia}, \xi^{(2)}_{kia} \in I(\tau_k,\tau_{k-1})$
such that
\begin{align}
  \bm{v}_i(\tau_{k-1})^\T
    &= \bm{v}_i(\tau_k)^\T + \frac{\Delta\tau}N \dot{\bm{v}}_i(\tau_k)^\T + \frac{\Delta\tau^2}{2N^2} \sum_{a=1}^n \bm{e}_a^\T \ddot{v}_{i,a}(\xi^{(1)}_{kia}),
    \vv\label{eq:proj-vector.utaylor.1} \\
  \bm{u}_i(\tau_k)
    &= \bm{u}_i(\tau_{k-1}) - \frac{\Delta\tau}N \dot{\bm{u}}_i(\tau_{k-1}) + \frac{\Delta\tau^2}{2N^2} \sum_{a=1}^n \bm{e}_a \ddot{u}_{i,a}(\xi^{(2)}_{kia}).
    \vv\label{eq:proj-vector.utaylor.2}
\end{align}
One also notices that
\begin{align}
  \dot P(\tauF) P(\tauF) &= -\sum_{i=1}^m \bm{u}_i(\tauF)\dot{\bm{v}}_i(\tauF)^\T P(\tauF).
  \vv\label{eq:proof.proj-derive-right-proj}
\end{align}
Substituting Eqs.~\eqref{eq:proj-vector.utaylor.1}--\eqref{eq:proof.proj-derive-right-proj}
into Eq.~\eqref{eq:proof.seq.deriv.3}, one obtains
\begin{align}
  \Df P_N(\tauF;\tauI)
    &= \dot P(\tauF) P_N(\tauF;\tauI) - \frac1N R_N(\tauF;\tauI),
    \vv\label{eq:proof.deriv-separate-residual}\\
  R_N(\tauF;\tauI)
    &\eqdef \frac{\Delta\tau}{N}
      \sum_{k=1}^{N-1} P_{k-1}(\tauF;\tau_{k-1})
      \sum_{i=1}^m \biggl\{
        \dot{\bm{u}}_i(\tau_{k-1})\dot{\bm{v}}_i(\tau_k)^\T \nonumber \\ & \quad
        + \frac{\Delta\tau}{2} \sum_{a=1}^n \left[
          (1-\tfrac{k-1}{N}) \dot{\bm{u}}_i(\tau_{k-1})\bm{e}_a^\T\ddot{v}_{i,a}(\xi^{(1)}_{kia})
          + (1-\tfrac{k}{N}) \bm{e}_a\ddot{u}_{i,a}(\xi^{(2)}_{kia})\dot{\bm{v}}_i(\tau_k)^\T\right]
      \biggr\} P_{N-k}(\tau_k;\tauI) \nonumber \\ & \quad
    + P_{N-1}(\tauF;\tau_{N-1}) \sum_{i=1}^m \dot{\bm{u}}_i(\tau_{N-1})
      \biggl[\frac{\Delta\tau}{N}\dot{\bm{v}}_i(\tauI)^\T
      + \frac{\Delta\tau^2}{2N^2}\sum_{a=1}^n \bm{e}_a^\T \ddot{v}_{i,a}(\xi_{Nia}^{(1)}) \biggr]
      P(\tauI).
      \vv\label{eq:proof.resseq}
\end{align}

\subsection{Boundedness of $P_N(\tauF;\tauI)$ and $R_N(\tauF;\tauI)$}
\newcommand{\tauK}{\tau_k}
\newcommand{\tauL}{\tau_l}

Here we will show the boundedness
of the sequences $\{P_N(\tauF;\tauI)\}_N$ and $\{R_N(\tauF;\tauI)\}_N$
in the limit $N\to\infty$.
More specifically we here find
upper bounds of those sequences which is independent of $N$.
First let us consider upper bounds of the single projector and its derivatives:
\begin{align}
  |P_{ab}(\tau)| &\le 1+mM_0^2, \\
  |\dot P_{ab}(\tau)| &\le 2mM_0 M_1, \\
  |\ddot P_{ab}(\tau)| &\le 2m(M_0M_2 + M_1^2),
\end{align}
where $P_{ab}(\tau) \eqdef \bm{e}_a^\T P(\tau) \bm{e}_b$
is the matrix element of the projector.
Matrix elements for sequences are similarly defined:
$P_{N,ab}(\tauF;\tauI) \eqdef \bm{e}_a^\T P_N(\tauF;\tauI)\bm{e}_b$,
and $R_{N,ab}(\tauF;\tauI) \eqdef \bm{e}_a^\T R_N(\tauF;\tauI)\bm{e}_b$.

From the mean-value theorem,
for any integers, $k$ and $l$, such that $0\le k < l \le N$,
there exists
$\xi^{(3)}_{kab} \in I(\tau_{k+1},\tau_k)$
such that
\begin{align}
  P_{l-k,ab}(\tauK;\tauL)
    &=\sum_{c=1}^n P_{ac}(\tauK) P_{l-k-1,cb}(\tau_{k+1};\tauL) \nonumber \\
    &=\sum_{c=1}^n \biggl[P_{ac}(\tau_{k+1})
        +\frac{\Delta\tau}{N} \dot P_{ac}(\xi^{(3)}_{kac})\biggr]
      P_{l-k-1,cb}(\tau_{k+1};\tauL) \nonumber \\
    &= P_{l-k-1,ab}(\tau_{k+1};\tauL)
      + \frac{\Delta\tau}{N} \sum_{c=1}^n \dot P_{ac}(\xi^{(3)}_{kac})
      P_{l-k-1,cb}(\tau_{k+1};\tauL).
\end{align}
Here one finds an upper bound as
\begin{align}
  \sup_{a,b}|P_{l-k,ab}(\tauK;\tauL)|
    &\le \left(1 + \frac{M'}{N}\right)
      \sup_{a,b}|P_{l-k-1,ab}(\tau_{k+1};\tauL)| \nonumber \\
    &\le \dots \le \left(1+\frac{M'}{N}\right)^{l-k} \sup_{a,b}|P_{ab}(\tauL)| \nonumber \\
    &\le e^{M'(l-k)/N} (1+mM_0^2),
      \vv\label{eq:proof.sup-partial-seqterm.1}
\end{align}
where $M' \eqdef 2|\Delta\tau|nmM_0M_1$.
In particular, when $k=0$ and $l=N$, one obtains $|P_{N,ab}(\tauF;\tauI)| \le e^{M'} (1+mM_0^2)$.
Using Eq.~\eqref{eq:proof.sup-partial-seqterm.1},
one also obtain an upper bound of $|R_N(\tauF;\tauI)|$ from the definition~\eqref{eq:proof.resseq} as
\begin{align}
  |R_{N,ab}(\tauF;\tauI)|
    &\le n^2 \frac{|\Delta\tau|}{N} \sum_{k=1}^{N-1} e^{M'(k-l)/N} (1+mM_0^2)
      m \biggl\{ M_1^2 \nonumber \\ & \quad
    + \frac{|\Delta\tau|}2 \left[(1-\tfrac{k-1}N)M_1M_2 + (1-\tfrac{k}N)M_1M_2\right]
      \biggr\}e^{M'(N-k)/N}(1+mM_0^2) \nonumber \\ & \quad
    + n^2 \frac{|\Delta\tau|}{N} e^{M'(N-1)/N}(1+mM_0^2)
      m M_1\left(M_1+\frac{|\Delta\tau|}{2N} M_2\right) (1+mM_0^2) \nonumber \\
    &= |\Delta\tau| e^{M'(N-1)/N} n^2m (1+mM_0^2)^2(M_1^2 + |\Delta\tau|M_1M_2/2) \nonumber \\
    &\le M_R(\Delta\tau) \eqdef |\Delta\tau| e^{M'} n^2m (1+mM_0^2)^2(M_1^2 + |\Delta\tau|M_1M_2/2).
\end{align}

\subsection{Compact convergence of $P_N(\tauF;\tauI)$}
Next we show that $\{P_N(\tauF;\tauI)\}_N$ is a Cauchy sequence.
For $N\ge2$, the difference between two consecutive sequence terms is
\begin{align}
  P_{N,ab}(\tauF;\tauI)-P_{N-1,ab}(\tauF;\tauI)
  &=[P_{N,ab}(\tauF;\tauI)-P_{N-1,ab}(\tau_1;\tauI)]
    -[P_{N-1,ab}(\tauF;\tauI)-P_{N-1,ab}(\tau_1;\tauI)] \nonumber \\
  &=\sum_{c=1}^n \left[\frac{\Delta\tau}{N}\dot P_{ac}(\tau_1)
      + \frac{\Delta\tau^2}{2N^2}\ddot P_{ac}(\xi^{(4)}_{0ac}) \right]
    P_{N-1,cb}(\tau_1;\tauI)
    \nonumber\\&\quad
    -\sum_{c=1}^n \frac{\Delta\tau}{N} \dot P_{ac}(\xi^{(5)}_{ab}) P_{N,cb}(\xi^{(5)}_{ab};\tauI)
    +\frac{\Delta\tau}{N(N-1)} R_{N-1,ab}(\xi^{(5)}_{ab};\tauI) \\
  &= -\frac{\Delta\tau}{N}(\xi^{(5)}_{ab}-\tau_1)
      \sum_{c=1}^n \D[\dot P_{ac}(\tau) P_{N-1,cb}(\tau;\tauI)]|_{\tau=\xi^{(6)}_{ab}}
    \nonumber\\&\quad
    +\sum_{c=1}^n
      \frac{\Delta\tau^2}{2N^2}\ddot P_{ac}(\xi^{(4)}_{0ac})
      P_{N-1,cb}(\tau_1;\tauI)
    +\frac{\Delta\tau}{N(N-1)} R_{N-1}(\xi^{(5)}_{ab};\tauI).
\end{align}
Here, the Taylor's theorem and the mean-value theorem were used:
there exist
$\xi^{(4)}_{0ab}, \xi^{(5)}_{ab} \in I(\tau_1,\tauF)$ and
$\xi^{(6)}_{ab}\in I(\tau_1,\xi^{(5)}_{ab})$
such that
\begin{align}
  P_{ab}(\tauF)
    &= P_{ab}(\tau_1)
      + \frac{\Delta\tau}{N} \dot P_{ab}(\tau_1)
      + \frac{\Delta\tau^2}{2N^2} \ddot P_{ab}(\xi^{(4)}_{0ab}), \\
  P_{N-1,ab}(\tauF;\tauI)
    &= P_{N-1,ab}(\tau_1;\tauI)
      + \frac{\Delta\tau}{N} \Df P_{N-1,ab}(\xi^{(5)}_{ab};\tauI), \\
  \sum_{c=1}^n \dot P_{ac}(\xi^{(5)}_{ab}) P_{N-1,cb}(\xi^{(5)}_{ab};\tauI)
    &= \sum_{c=1}^n \dot P_{ac}(\tau_1) P_{N-1,cb}(\tau_1;\tauI)
    \nonumber\\&\quad
      + (\xi^{(5)}_{ab}-\tau_1) \sum_{c=1}^n \D [ \dot P_{ac}(\tau) P_{N-1,cb}(\tau;\tauI) ]|_{\tau=\xi^{(6)}_{ab}}.
\end{align}
Then one finds an upper bound of the difference of the two consecutive terms as
\begin{align}
  &\hspace{-1em}
    \biggl|\sum_{c=1}^n \D [ \dot P_{ac}(\tau) P_{N-1,cb}(\tau;\tauI) ]\biggr| \nonumber\\
    &=\left| \sum_{d=1}^n \biggl[
        \ddot P_{ac}(\tau)
        +\sum_{c=1}^n \dot P_{ac}(\tau) \dot P_{cd}(\tau)
      \biggr] P_{N-1,db}(\tau;\tauI)
      -\sum_{c=1}^n \frac1{N-1}\dot P_{ac}(\tau) R_{N-1,cb}(\tau;\tauI) \right| \nonumber\\
    &\le n [2m(M_0M_2 + M_1^2) + n(2mM_0M_1)^2] (1+mM_0^2) e^{M'}
      +n \cdot \frac1{N-1} 2mM_0M_1 M_R(\tau-\tauI) \nonumber\\
    &\le 2mn[ M_0M_2 + M_1^2 + 2n m M_0^2M_1^2] (1+mM_0^2) e^{M'}
      +2mnM_0M_1 M_R(\Delta\tau) =: M^{(1)}(\Delta\tau), \\
  &\hspace{-1em}
    |P_{N,ab}(\tauF;\tauI)-P_{N-1,ab}(\tauF;\tauI)| \nonumber\\
    &\le \frac{|\Delta\tau(\xi^{(5)}_{ab}-\tau_1)|}{N} M^{(1)}(\Delta\tau)
      +n\cdot \frac{|\Delta\tau|^2}{2N^2} \cdot2m(M_0M_2+M_1^2) (1+mM_0^2) e^{M'}
      +\frac{|\Delta\tau|}{N(N-1)} M_R(\xi^{(5)}_{ab}-\tauI) \nonumber\\
    &\le \frac1{N(N-1)}\left[
      |\Delta\tau|^2 M^{(1)}(\Delta\tau)
      +n\cdot |\Delta\tau|^2 \cdot m(M_0M_2+M_1^2) (1+mM_0^2) e^{M'}
      +|\Delta\tau| M_R(\Delta\tau)\right] \nonumber\\
    & =: \frac{1}{N(N-1)}M^{(2)}(\Delta\tau).
    \vv\label{eq:proof.difference-upper-bound}
\end{align}
To obtain the second inequality in Eq.~\eqref{eq:proof.difference-upper-bound},
we used $|\xi^{(5)}_{ab}-\tau_1| \le |\Delta\tau|/N$, $1/N \le 1/(N-1)$
and the fact that $M_R(\Delta\tau)$ is monotonically increasing with respect to $|\Delta\tau|$.

Now we are ready to show that the sequence is a Cauchy sequence.
For integers $N_1$, $N_2$ such that $N_2>N_1\ge 1$,
\begin{align}
  0 &\le |P_{N_2,ab}(\tauF;\tauI)-P_{N_1,ab}(\tauF;\tauI)| \nonumber \\
    &\le \sum_{N=N_1+1}^{N_2} |P_{N,ab}(\tauF;\tauI)-P_{N-1,ab}(\tauF;\tauI)| \nonumber \\
    &\le \sum_{N=N_1+1}^{N_2} \frac1{N(N-1)} M^{(2)}(\Delta\tau)
    = \left(\frac1{N_1}-\frac1{N_2}\right) M^{(2)}(\Delta\tau).
    \vv\label{eq:proof.cauchy-upper-bound}
\end{align}
Therefore
the sequence is a Cauchy sequence and pointwise convergent:
\begin{align}
  \lim_{\begin{subarray}{c}N_2\to\infty \\ N_1\to\infty\end{subarray}}
  |P_{N_2,ab}(\tauF;\tauI)-P_{N_1,ab}(\tauF;\tauI)| = 0,
\end{align}
Then a pathline projector can be defined
as the limit value of the sequence:
\begin{align}
  P(\tauF;\tauI) &\eqdef \lim_{N\to\infty} P_N(\tauF;\tauI),
\end{align}
which ensures that the limits in Eqs.~\eqref{eq:pathline.def3} and \eqref{eq:pathline.def5} are convergent,
and the original pathline projectors, $\Delta(\tauF;\tauI)\ud{\mu}{\alpha}$ and $\Delta(\tauF;\tauI)\ud{\mu\nu}{\alpha\beta}$,
are well-defined.
Taking the limit $N\to\infty$ for Eqs.~\eqref{eq:proof.seqprop.1}--\eqref{eq:proof.seqprop.3},
one obtains the following relations for the pathline projectors:
\begin{align}
  P(\tauF;\tauI) &= P(\tauF)P(\tauF;\tauI) = P(\tauF;\tauI)P(\tauI),
    \vv\label{eq:proof.prop.1} \\
  P(\tauI;\tauI) &= P(\tauI),
    \vv\label{eq:proof.prop.2} \\
  P(\tauF;\tauI) &= P(\tauI;\tauF)^\T,
    \vv\label{eq:proof.prop.3}
\end{align}
which are equivalent to the properties~\eqref{eq:ttproj3.prop-repr}--\eqref{eq:ttproj5.prop-sym}
of the pathline projectors, $\Delta(\tauF;\tauI)\ud{\mu}{\alpha}$ and $\Delta(\tauF;\tauI)\ud{\mu\nu}{\alpha\beta}$.
Also, taking the limit $N_2\to\infty$ for Eq.~\eqref{eq:proof.cauchy-upper-bound}, it follows that
\begin{align}
  |P_{ab}(\tauF;\tauI) - P_{N_1,ab}(\tauF;\tauI)|
    &\le \frac{M^{(2)}(\Delta\tau)}{N_1}.
    \vv\label{eq:proof.difference-limit-seq}
\end{align}
\newcommand{\tauA}{\tau_\mathrm{A}}
\newcommand{\tauB}{\tau_\mathrm{B}}

It can even be shown that the sequence is compactly convergent.
For any compact closed interval $[\tauA,\tauB] \subset D$,
and for any $\tauF, \tauI \in [\tauA,\tauB]$,
\begin{align}
  0 \le |P_{N,ab}(\tauF;\tauI) - P_{ab}(\tauF;\tauI)|
    &\le \frac{M^{(2)}(\Delta\tau)}{N} \le \frac{M^{(2)}(|\tauB-\tauA|)}{N}.
\end{align}
Therefore the sequence is uniformly convergent
in the compact interval $[\tauA,\tauB]$:
\begin{align}
  \lim_{N\to\infty}\sup_{\tauF,\tauI\in[\tauA,\tauB]}
    |P_{N,ab}(\tauF;\tauI) - P_{ab}(\tauF;\tauI)|
    &= 0,
\end{align}
which means that the sequence is compactly convergent in $D$.

\subsection{Compact convergence of $\Df P_N(\tauF;\tauI)$}
For any compact closed interval $[\tauA,\tauB] \subset D$,
and for any $\tauF, \tauI \in [\tauA,\tauB]$,
\begin{align}
  0 &\le \biggl|\Df P_{N,ab}(\tauF;\tauI)
      - \sum_{c=1}^n \dot P_{ac}(\tauF) P_{cb}(\tauF;\tauI)\biggr| \nonumber \\
    &\le\frac{1}{N} |R_{N,ab}(\tauF;\tauI)|
      +\sum_{c=1}^n |\dot P_{ac}(\tauF)|\cdot|P_{N,cb}(\tauF;\tauI) - P_{cb}(\tauF;\tauI)| \nonumber \\
    &\le\frac{1}{N} [ M_R(\Delta\tau)+ 2nmM_0M_1 M^{(2)}(\Delta\tau)] \nonumber \\
    &\le\frac{1}{N} [ M_R(\tauB-\tauA)+ 2nmM_0M_1 M^{(2)}(\tauB-\tauA)]
      =: \frac 1N M^{(3)}(\tauB-\tauA).
\end{align}
Here one can use Eq.~\eqref{eq:proof.deriv-separate-residual} to obtain the second line
and Eq.~\eqref{eq:proof.difference-limit-seq} to obtain the third line.
To obtain the fourth line, one can use the fact that
$M_R(\Delta\tau)$ and $M^{(2)}(\Delta\tau)$ are monotonically increasing with $|\Delta\tau|$.
Therefore the derivative of the sequence is uniformly convergent in the closed interval $[\tauA,\tauB]$:
\begin{align}
  \lim_{N\to\infty} \sup_{\tauF,\tauI\in [\tauA,\tauB]}
    \biggl|\Df P_{N,ab}(\tauF;\tauI)
      - \sum_{c=1}^n \dot P_{ac}(\tauF) P_{cb}(\tauF;\tauI)\biggr| &= 0,
\end{align}
which means that the derivative $\Df P_N(\tauF;\tauI)$
is compactly convergent in the domain $D$.
Thus the limit and the derivative commute:
\begin{align}
  \Df P(\tauF;\tauI)
    &= \Df \lim_{N\to\infty }P_N(\tauF;\tauI)
    = \lim_{N\to\infty} \Df P_N(\tauF;\tauI)
    = \Df P(\tauF) P(\tauF;\tauI).
\end{align}
Here, the properties \eqref{eq:ttproj3.prop-dyn} and \eqref{eq:ttproj5.prop-dyn} are obtained.

Finally, let us consider the derivative of the product of two pathline projectors:
\begin{align}
  \D[P(\tauF;\tau) P(\tau;\tauI)]
    &= 2 P(\tauF;\tau) \dot P(\tau) P(\tau;\tauI) \nonumber \\
    &= -\sum_{i=1}^m 2 P(\tauF;\tau)
      [\dot{\bm{u}}_i(\tau)\bm{v}_i(\tau)^\T + \bm{u}_i(\tau)\dot{\bm{v}}_i(\tau)^\T]
      P(\tau;\tauI) \nonumber \\
    &= 0,
\end{align}
where one can use Eq.~\eqref{eq:proof.single-derivative.1} to obtain the second line
and $P(\tau)\bm{u}_i(\tau) = \bm{v}_i(\tau)^\T P(\tau) = 0$ to obtain the third line.
Thus, the expression
$P(\tauF;\tau) P(\tau;\tauI)$ is unchanged for the change of $\tau$.
Choosing $\tauF$ as $\tau$, one obtains
\begin{align}
  P(\tauF;\tau) P(\tau;\tauI)
    &= P(\tauF;\tauF) P(\tauF;\tauI) \nonumber \\
    &= P(\tauF;\tauI),
\end{align}
which is equivalent to the properties
\eqref{eq:ttproj3.prop-connect} and \eqref{eq:ttproj5.prop-connect}.
Here one can use Eqs.~\eqref{eq:proof.prop.2} and \eqref{eq:proof.prop.1} to obtain the second line.

It should be noted that Eq.~\eqref{eq:ttproj5.prop-dyn2} is obtained from
Eq.~\eqref{eq:ttproj5.prop-dyn} by a property
of the spatial traceless symmetric projector $\Delta\ud{\mu\nu}{\alpha\beta}$ as follows:
\begin{align}
  \D\Delta\ud{\mu\nu}{\kappa\lambda} \Delta\ud{\kappa\lambda}{\alpha\beta}
  &= \frac12(\D\Delta\ud\mu\kappa \Delta\ud\nu\lambda
    + \Delta\ud\mu\kappa \D\Delta\ud\nu\lambda
    + \D\Delta\ud\mu\lambda \Delta\ud\nu\kappa
    + \Delta\ud\mu\lambda \D\Delta\ud\nu\kappa)\Delta\ud{\kappa\lambda}{\alpha\beta} \nonumber \\ & \quad
    + \frac13(\D\Delta^{\mu\nu}\Delta_{\kappa\lambda}
    + \Delta^{\mu\nu}\D\Delta_{\kappa\lambda})\Delta\ud{\kappa\lambda}{\alpha\beta} \nonumber \\
  &= \frac12(\D\Delta\ud\mu\kappa g\ud\nu\lambda
    + g\ud\mu\kappa \D\Delta\ud\nu\lambda
    + \D\Delta\ud\mu\lambda g\ud\nu\kappa
    + g\ud\mu\lambda \D\Delta\ud\nu\kappa
    )\Delta\ud{\kappa\lambda}{\alpha\beta} \nonumber \\ & \quad
    + \Delta^{\mu\nu}(-\D u_\kappa u_\lambda - u_\kappa \D u_\lambda) \Delta\ud{\kappa\lambda}{\alpha\beta} \nonumber \\
  &= (\D\Delta\ud\mu\kappa g\ud\nu\lambda
    + g\ud\mu\kappa \D\Delta\ud\nu\lambda)\Delta\ud{\kappa\lambda}{\alpha\beta},
\end{align}
where one can use Eqs.~\eqref{eq:pathline.proj-prop-traceless} and \eqref{eq:pathline.proj-prop-spatial}
to obtain the second line
and Eqs.~\eqref{eq:pathline.proj-prop-sym5} and \eqref{eq:pathline.proj-prop-trans5} to obtain the third line.

%%%%%%%%%%%%%%%%%%%%%%%%%%%%%%%%%%%%%%%%%%%%%%%%%%%%%%%%%%%%%%%%%%%%%%%%%%%%%%%
\section{Right and left eigenvectors of kernels of projectors}
\label{sec:pathline.appendix}

Here we explicitly construct $\{\bm{u}_i\}_{i=1}^m$ and $\{\bm{v}_i^\T\}_{i=1}^m$ in Appendix~\ref{sec:pathline.proof}
for the projectors, $\Delta\ud\mu\alpha$ and $\Delta\ud{\mu\nu}{\alpha\beta}$.
For the spatial projector $\Delta\ud\mu\alpha$, the dimension of the kernel is $m=1$,
and the eigenvectors can be trivially defined as
\begin{align}
  u_{1,\alpha} &= u^\alpha, & v_{1,\alpha} &= u_\alpha,
\end{align}
where $u_{i,\alpha}$ and $v_{i,\alpha}$ are the $\alpha$-th components of the $\bm{u}_i$ and $\bm{v}_i^\T$.

For the spatial traceless symmetric projector $\Delta\ud{\mu\nu}{\alpha\beta}$,
the dimension of the kernel is $m=11$.
It is nontrivial to construct the set of the eigenvectors by $u^\mu$
because there is freedom degrees which is not entirely fixed by $u^\mu$,
\thatis, freedom degrees of the rotation in the local rest frame for spatial indices.
We here fix the rotation degrees of freedom
by introducing a specific choice of the basis $\langle a_0^\mu, a_1^\mu, a_2^\mu, a_3^\mu\rangle$:
\begin{align}
  \begin{pmatrix}
    \bm{a}_0 & \bm{a}_1 & \bm{a}_2 & \bm{a}_3
  \end{pmatrix} &\eqdef
  \begin{pmatrix}
    u^0 & \bm{u}^\T \\
    \bm{u} & 1 + \frac{\bm{u}\bm{u}^\T}{u^0 + 1}
  \end{pmatrix},
\end{align}
where $(u^0, \bm{u}) \eqdef u^\mu$.
The right-hand side is actually
the representation
of the Lorentz transformation, $\Lambda\ud\mu\alpha$,
of the boost by $u^\mu$.
The vectors $a_i^\mu$ ($i=1,2,3$), which are written in terms of $u^\mu$,
introduce specific directions of the spatial components.
Corresponding covariant vectors $\langle b_{0\mu}, b_{1\mu}, b_{2\mu}, b_{3\mu}\rangle$ are defined as
\begin{align}
  \begin{pmatrix}
    \bm{b}_0^\T \\ \bm{b}_1^\T \\ \bm{b}_2^\T \\ \bm{b}_3^\T
  \end{pmatrix} &\eqdef
  \begin{pmatrix}
    \bm{a}_0 & \bm{a}_1 & \bm{a}_2 & \bm{a}_3
  \end{pmatrix}^{-1} =
  \begin{pmatrix}
    u^0 & -\bm{u}^\T \\
    -\bm{u} & 1 + \frac{\bm{u}\bm{u}^\T}{u^0 + 1}
  \end{pmatrix}.
\end{align}
We note that $a_0^\mu = u^\mu$ and $b_{0\mu} = u_\mu$.
With these bases, $\{a_i^\mu\}$ and $\{b_{i\mu}\}$,
the right and left eigenvectors can be constructed as follows:
The eigenvectors for the temporal--temporal tensor component are
\begin{align}
%% temporal-temporal
  u_{1,\mu\nu} &= u^\mu u^\nu,   & v_{1,\mu\nu} &= u_\mu u_\nu.
\end{align}
The eigenvectors for the temporal--spatial tensor components are
\begin{align}
%% temporal-spatial
  u_{2,\mu\nu} &= u^\mu a_1^\nu, & v_{2,\mu\nu} &= u_\mu b_{1\nu}, \\
  u_{3,\mu\nu} &= u^\mu a_2^\nu, & v_{3,\mu\nu} &= u_\mu b_{2\nu}, \\
  u_{4,\mu\nu} &= u^\mu a_3^\nu, & v_{4,\mu\nu} &= u_\mu b_{3\nu}, \\
  u_{5,\mu\nu} &= a_1^\nu u^\mu, & v_{5,\mu\nu} &= b_{1\mu} u_\nu, \\
  u_{6,\mu\nu} &= a_2^\nu u^\mu, & v_{6,\mu\nu} &= b_{2\mu} u_\nu, \\
  u_{7,\mu\nu} &= a_3^\nu u^\mu, & v_{7,\mu\nu} &= b_{3\mu} u_\nu.
\end{align}
The eigenvectors for the spatial antisymmetric components are
\begin{align}
%% spatial antisymmetric
  u_{8,\mu\nu}  &= a_1^\mu a_2^\nu - a_2^\mu a_1^\nu, & v_{8,\mu\nu}  &= b_{1\mu} b_{2\nu} - b_{2\mu} b_{1\nu}, \\
  u_{9,\mu\nu}  &= a_2^\mu a_3^\nu - a_3^\mu a_2^\nu, & v_{9,\mu\nu}  &= b_{2\mu} b_{3\nu} - b_{3\mu} b_{2\nu}, \\
  u_{10,\mu\nu} &= a_3^\mu a_1^\nu - a_1^\mu a_3^\nu, & v_{10,\mu\nu} &= b_{3\mu} b_{1\nu} - b_{1\mu} b_{3\nu}.
\end{align}
The eigenvectors for the spatial trace component are
\begin{align}
%% trace
  u_{11,\mu\nu} &= \Delta^{\mu\nu}, & v_{11,\mu\nu} &= \frac13\Delta_{\mu\nu}.
\end{align}
We note that all of these vectors, $u_{i,\mu\nu}(\tau)$ and $v_{i,\mu\nu}(\tau)$,
are $C^2$ functions of $\tau$ provided that $u^\mu(\tau)$ is a $C^2$ function
because all of these vectors are regular function of $u^\mu$.

%%%%%%%%%%%%%%%%%%%%%%%%%%%%%%%%%%%%%%%%%%%%%%%%%%%%%%%%%%%%%%%%%%%%%%%%%%%%%%
\section{Proof of restrictions on differential form: Eqs.~\eqref{eq:fdr.white.polynomial-L}--%
\eqref{eq:fdr.white.power-spectrum}}
\label{sec:white}

Here we show Eqs.~\eqref{eq:fdr.white.polynomial-L}--\eqref{eq:fdr.white.power-spectrum}
for a single-component dissipative current $\Gamma$.

\subsection{Restriction by positive semi-definiteness of FDR}
First we consider the positive semi-definiteness of the autocorrelation
$\langle\xi(x)\xi(x')\rangle$ where its spatial dependence is seen as matrix indices.
Reflecting the translational symmetry of the equilibrium systems,
the autocorrelation is diagonalized in the Fourier space as in Eq.~\eqref{eq:chf.general-differential-noise-fdr}.
The condition for the positive semi-definiteness can be obtained from Eq.~\eqref{eq:chf.general-differential-noise-fdr-power}:
\begin{align}
  I_{\omega,\bm{k}} = 2T\kappa \Re [M(-i\bm{k})L(-i\omega,-i\bm{k})] \ge 0,
  \vv\label{eq:white.positive-semidefinite}
\end{align}
where we used the fact that $M(-i\bm{k}) = M(-i\bm{k})^*$ because
$M(\bm{w})$ is an even polynomial \eqref{eq:chf.general-polynomial-even-M} with real-valued coefficients
so that $M(-i\bm{k}) = M_2((-i\bm{k})\otimes(-i\bm{k})) = M_2(-\bm{k}\otimes\bm{k}) \in \mathbb{R}$.

The polynomial $L(-i\omega,-i\bm{k})$ is factorized with respect to $-i\omega$ as
\begin{align}
  L(-i\omega,-i\bm{k})
   &= (-1)^N A_{\bm{k}} \prod_{p=1}^N i(\omega - \omega_p(\bm{k})),
\end{align}
where $N = \deg_\omega L$
is the degree of $-i\omega$ in the polynomial $L(-i\omega,-i\bm{k})$.
The factor $A_{\bm{k}}$ is the coefficient of the highest-order term $A_{\bm{k}}(-i\omega)^N$.
The factor $A_{\bm{k}}$ is real because $L(-i\omega,-i\bm{k})$ is
an even polynomial of $-i\bm{k}$~\eqref{eq:chf.general-polynomial-even-L} with real-valued coefficients.
Here a wave number $\bm{k}$ is chosen so that $M(-i\bm{k})A_{\bm{k}}\ne 0$
because the positive semi-definiteness is trivially fulfilled when $M(-i\bm{k})A_{\bm{k}} = 0$.
The zeroes $\{\omega_p(\bm{k})\}_{p=1}^N$ of the polynomial $L(-i\omega,-i\bm{k})$
correspond to the poles of the memory function
$G_{\omega,\bm{k}} = M(-i\bm{k})/L(-i\omega,-i\bm{k})$.
Note that none of the zeroes $\{\omega_p(\bm{k})\}_{p=1}^N$ are canceled with a zero of $M(-i\bm{k})$
because $M(-i\bm{k})$ and $L(-i\omega,-i\bm{k})$ are defined to be coprime.
To ensure
the retardation~\eqref{eq:hydro.memory-function.retardation}
and the relaxation~\eqref{eq:hydro.memory-function.relaxation}
of the memory function $G(x-x')$,
the imaginary part of the poles should be negative:
\begin{align}
  \Im \omega_p(\bm{k}) &< 0, \quad(1\le p \le N).
  \vv\label{eq:white.positive-semidefinite.pole}
\end{align}

Here let us consider the complex argument of the polynomials:
\begin{align}
  \arg [M(-i\bm{k})L(-i\omega,-i\bm{k})]
    &= \arg[M(-i\bm{k}) A_{\bm{k}}] + \sum_{p=1}^N \arg[\omega-\omega_p(\bm{k})] -\frac{N\pi}2.
\end{align}
When the frequency $\omega$ goes from the negative infinity to the positive infinity,
the argument of the factor $\omega - \omega_p(\bm{k})$
is continuously changed from $0$ to $-\pi$ because $\Im\omega_p(\bm{k})< 0$.
Thus the whole argument continuously decreases by $N\pi$:
\begin{align}
  \lim_{\omega\to\infty} \arg[M(-i\bm{k})L(-i\omega,-i\bm{k})]
  -\lim_{\omega\to-\infty} \arg[M(-i\bm{k})L(-i\omega,-i\bm{k})]
    &= -N\pi.
  \vv\label{eq:white.positive-semidefinite.arg-limit}
\end{align}

The positive semi-definiteness condition
\eqref{eq:white.positive-semidefinite}
is expressed in terms of the complex argument as
\begin{align}
  \arg[M(-i\bm{k})L(-i\omega,-i\bm{k})]
    &\in \left(2m\pi -\frac{\pi}2,2m\pi + \frac{\pi}2\right),
\end{align}
where $m$ is an integer.
Here the complex argument cannot be continuously changed by more than $\pi$,
and therefore the degree of the polynomial should be $N=0$ or $1$.

Now the polynomial $L(-i\omega,-i\bm{k})$ has the form:
\begin{align}
  L(-i\omega,-i\bm{k}) &= -i\omega A_{\bm{k}} + B_{\bm{k}},
  \vv\label{eq:white.positivie-semidefinite.restricted-L}
\end{align}
where $A_{\bm{k}}$ and $B_{\bm{k}}$
are even polynomials of $\bm{k}$
and real due to the property of the polynomial $L(-i\omega,-i\bm{k})$~\eqref{eq:chf.general-polynomial-even-L}.
The $N=0$ case corresponds to the case $A_{\bm{k}}=0$.
For the case $N=1$, $B_{\bm{k}}$ should have the same sign with $A_{\bm{k}}$
because the imaginary part of the pole $\omega_1 = -iB_{\bm{k}}/A_{\bm{k}}$ should be negative~\eqref{eq:white.positive-semidefinite.pole}.
Also, $M(-i\bm{k})$ should have the same sign with $A_{\bm{k}}$ and $B_{\bm{k}}$ unless $M(-i\bm{k}) = 0$
so that the positive semi-definiteness is satisfied.

\subsection{Restriction by causality in memory function}
Another restriction comes from the causality of the memory function.
The general conditions that the response function
in the derivative expansion respects the causality are given in Ref.~\cite{Minami:2013aia}:
\begin{itemize}
\item
  The coefficient of the highest order of $\omega$ does not depend on $\bm{k}$.
\item
  The poles $\{\omega_p(\bm{k})\}_{p=1}^N$ of the memory function
  should satisfy the following equations:
  \begin{align}
    \lim_{|\bm{k}|\to\infty} \left|\frac{\Re \omega_p(\bm{k})}{|\bm{k}|}\right| &< 1,
    \vv\label{eq:white.causality.pole-real} \\
    \lim_{|\bm{k}|\to\infty} \left|\frac{\Im \omega_p(\bm{k})}{|\bm{k}|}\right| &< \infty.
    \vv\label{eq:white.causality.pole-imag}
  \end{align}
\end{itemize}

Let us apply these conditions to
the polynomial $L(-i\omega,-i\bm{k})$~\eqref{eq:white.positivie-semidefinite.restricted-L}.
Because the highest-order coefficients are constants,
$B_{\bm{k}} = B$ for the case $N=0$, and $A_{\bm{k}} = A$ for the case $N=1$.
For the case $N=1$, the pole $\omega_1(\bm{k}) = -iB_{\bm{k}}/A$ is pure imaginary,
and thus Eq.~\eqref{eq:white.causality.pole-real} is already satisfied.
To satisfy Eq.~\eqref{eq:white.causality.pole-imag},
the degree of $\bm{k}$ in $B_{\bm{k}}$ should not be larger than one.
Since $B_{\bm{k}}$ is an even polynomial of $\bm{k}$,
$B_{\bm{k}}$ is a constant also in the case $N=1$.
Because of the normalization~\eqref{eq:chf.general-polynomial-one-L},
$B=1$ in both cases of $N$.
In the case $N=1$, $A$ is positive because its sign is the same as $B=1$,
and in fact $A$ physically corresponds to the relaxation time $A = \tau_R > 0$.
Therefore one obtains the following form of the polynomial $L$:
\begin{align}
  L(-i\omega, -i\bm{k}) = 1 - i\omega\tau_R,
  \vv\label{eq:white.causality.restricted-L}
\end{align}
which is equivalent to \eqref{eq:fdr.white.polynomial-L}.
It should be noted that the $N=0$ case corresponds to the case $\tau_R=0$.
For both cases of $N$, also $M(-i\bm{k})$ should be non-negative
because its sign is the same as $B=1$.
Also Eq.~\eqref{eq:fdr.white.polynomial-M} is shown
because $M(-i\bm{k})$ should have the same sign with $B=1$ unless $M(-i\bm{k}) = 0$.
Eq.~\eqref{eq:fdr.white.power-spectrum} is immediately obtained
by substituting \eqref{eq:white.causality.restricted-L} into \eqref{eq:white.positive-semidefinite}.

%%%%%%%%%%%%%%%%%%%%%%%%%%%%%%%%%%%%%%%%%%%%%%%%%%%%%%%%%%%%%%%%%%%%%%%%%%%%%%

\bibliographystyle{apsrev4-1}
\bibliography{paper-rfh2}
\end{document}